\documentclass[12pt]{iopart}
\usepackage{graphicx,dsfont}

\newcommand{\bra}[1]{\langle #1 \vert}
\newcommand{\ket}[1]{\vert #1 \rangle}
\newcommand{\braket}[2]{\langle #1 \vert #2 \rangle}
\newcommand{\vev}[1]{\langle #1 \rangle}

\newcommand{\tUU}{\tau^{\scriptscriptstyle UU}}

\newcommand{\tLL}{\tau^{\scriptscriptstyle LL}}

\newcommand{\betaB}{\beta^{\scriptscriptstyle (\textrm{\tiny B})}}

\newcommand{\overleftrightarrow}[1]{\stackrel{\longleftrightarrow}{#1}}
\newcommand{\xrightarrow}[1]{\stackrel{#1}{\longrightarrow}}

\begin{document}
\title{Goodness-of-fit tests with dependent observations}
\author{R\'emy Chicheportiche$^{1,2}$ and Jean-Philippe Bouchaud$^1$}
\address{$^1$ Capital~Fund~Management, 6--8 boulevard Haussmann, 75\,009 Paris, France}
\address{$^2$ Chair~of~Quantitative~Finance, Laboratory of Applied Mathematics and Systems,
Ecole~Centrale~Paris, 92\,290 Ch\^atenay-Malabry, France
}
\eads{\mailto{remy.chicheportiche@ecp.fr}, \mailto{jean-philippe.bouchaud@cfm.fr}}
\date{\today}

\begin{abstract}
We revisit the Kolmogorov-Smirnov and Cram\'er-von Mises goodness-of-fit (GoF) tests 
and propose a generalisation to identically distributed, but dependent univariate random variables.
We show that the dependence leads to a reduction of the ``effective'' number of independent observations. 
The generalised GoF tests are not distribution-free but rather depend on all the lagged bivariate copulas.
These objects, that we call ``self-copulas'', encode all the non-linear temporal dependences. We introduce a specific, 
log-normal model for these self-copulas, for which a number of analytical results are derived. An application to 
financial time series is provided. As is well known, the dependence is to be long-ranged in this case, a finding that
we confirm using self-copulas. As a consequence, the acceptance rates for GoF tests are substantially higher than if the
returns were iid random variables.\\[20pt]
\end{abstract}

\maketitle


\section{Introduction}

Goodness-of-Fit (GoF) tests are designed to assess quantitatively whether a sample of $N$ observations
can statistically be seen as a collection of $N$ \emph{independent} realizations of a given probability law, 
or whether two such samples are drawn from the same hypothetical distribution.
Two well-known and broadly used tests in this class are the Kolmogorov-Smirnov (KS) and Cram\' er-von Mises (CM) tests,
that both quantify how close an empirical cumulative distribution function (cdf) $F_N$ is from a target theoretical cdf $F$ 
(or from another empirical cdf) --- see Ref.~\cite{darling1957kolmogorov} for a nice review and many references.
The major strength of these tests lies in the fact that the asymptotic distributions of their test statistics 
is completely independent of the null-hypothesis cdf.

It  however so happens that in certain fields (physics, finance, geology, etc.) the random variable under scrutiny has some memory. 
Whereas the unconditional law of the variable may well be unique and independent of time, 
the conditional probability distribution of an observation \emph{following} a previous observation 
exhibits specific patterns, and in particular long-memory, even when the linear correlation is short-ranged or trivial. 
Examples of such phenomena can be encountered in fluid mechanics (the velocity of a turbulent fluid) 
and finance (stock returns have small auto-correlations but exhibit strong volatility clustering, a form of heteroscedasticity).
The long-memory nature of the underlying processes makes it inappropriate to use standard GoF tests in these cases. 
Still, the determination of the unconditional distribution of returns is a classic problem in quantitative finance, with obvious 
applications to risk control, portfolio optimization or derivative pricing. Correspondingly, the distribution of stock returns 
(in particular the behaviour of its tails) has been the subject of numerous empirical and theoretical papers (see e.g. \cite{plerou1999scaling,dragulescu2002probability} and
for reviews \cite{bouchaud2003theory,malevergne2006extreme} and references therein). Clearly, precise 
statements are only possible if meaningful GoF tests are available.

As a tool to study the --- possibly highly non-linear --- correlations between returns, ``copulas'' have long been used in actuarial 
sciences and finance to describe and model cross-dependences of assets, often in a risk management perspective \cite{embrechts2003modelling_art,embrechts2002correlation_art,malevergne2006extreme}.
Although the widespread use of simple analytical copulas to model multivariate dependences is more and more criticized \cite{mikosch2006copulas,chicheportiche2010joint}, 
copulas remain useful as a tool to investigate empirical properties of multivariate data \cite{chicheportiche2010joint}.

More recently, copulas have also been studied in the context of auto-dependent univariate time series, 
where they find yet another application range: just as Pearson's $\rho$ coefficient is commonly used to 
measure both linear cross-dependences and temporal correlations, 
copulas are well-designed to assess non-linear dependences both across assets or in time 
\cite{beare2010copulas,ibragimov2008copulas,patton2009copula_art} --- we will speak of ``self-copulas'' in the latter case. 
Interestingly, when trying to extend GoF tests to dependent variables, self-copulas appear naturally. 
In our empirical study of financial self-copulas, we rely on a non-parametric estimation rather than imposing, 
for example, a Markovian structure of the underlying process, as in e.g.~\cite{darsow1992copulas,ibragimov2008copulas}.

The organisation of the paper is in three parts. 
In Section~\ref{sec:GoF} we study theoretically how to account for general dependence in GoF tests: 
we first describe the statistical properties of the empirical cdf of a non-iid vector of observations of finite size,
as well as measures of its difference with an hypothesized cdf. We then study the limit properties of 
this difference and the asymptotic distributions of two norms.
In Section~\ref{sec:example} we go through a detailed example when the dependences are weak and described by a pseudo-elliptical copula.
Section~\ref{sec:application} is dedicated to an application of the theory to the case of financial data:
after defining our data set, we perform an empirical study of dependences in series of stock returns, 
and interpret the results in terms of the ``self-copula'';
implication of the dependences on GoF tests are illustrated for this special case using Monte-Carlo simulations.
The concluding section summarizes the main ideas of the paper, 
and technical calculations of sections \ref{sec:GoF} and \ref{sec:example} are collected in the appendix.

\section{Goodness-of-fit tests for a sample of dependent draws}\label{sec:GoF}

\subsection{Empirical cumulative distribution and its fluctuations}
Let $X$ be a latent random vector with $N$ identically distributed but dependent variables, with marginal cdf $F$. 
One realization of $X$ consists of a time series $\{x_1,\ldots,x_n, \ldots, x_N\}$ that exhibits some sort of persistence.
For a given number $x$ in the support of $F$, 
let $Y(x)$ be the random vector the components of which are the Bernoulli variables $Y_n(x)=\mathds{1}_{\{X_n\leq x\}}$.
The expectation value and the covariance of $Y_n(x)$ are given by:
\begin{eqnarray}
	\mathds{E}[Y_n(x)]=F(x),\\
	{\rm Cov}(Y_n(x),Y_m(x'))=F_{nm}(x,x')=C_{nm}(F(x),F(x')),
\end{eqnarray}
where by definition $C_{nm}$ is the ``copula'' of the random pair $(X_n,X_m)$.
The centered mean of $Y(x)$ is: 
\begin{equation}
	\overline{Y}(x)=\frac{1}{N}\sum_{n=1}^NY_n(x)-F(x)=\vev{Y_n(x)}_n-F(x)
\end{equation}
which measures the difference between the empirically determined cumulative distribution function at point $x$ and its true value.
It is therefore the quantity on which any statistics for Goodness-of-Fit testing is built.
Denoting $u=F(x),v=F(x')$, the covariance function of $\overline{Y}$ is easily shown to be:
\begin{equation}\label{eq:CovYbar}
	{\rm Cov}(\overline{Y}(u),\overline{Y}(v))=\frac{1}{N}\big(\min(u,v)-uv\big) \left[1 + \Psi_N(u,v)\right]
\end{equation}
where
\begin{equation}\label{eq:Psi}
	\Psi_N(u,v)= \frac{1}{N} \sum_{n, m \neq n}^N \frac{C_{nm}(u,v)-uv}{\min(u,v)-uv}
\end{equation}
measures the departure from the independent case, corresponding to $C_{nm}(u,v)=uv$ (in which case $\Psi_N(u,v) \equiv 0$). 
Note that decorrelated but dependent variables may lead to a non zero value of $\Psi_N$, 
since the whole pairwise copula enters the formula and not only the linear correlation coefficients.
When the denominator is zero, the fraction should be understood in a limit sense;
we recall in particular that \cite{chicheportiche2010joint}
\begin{equation}
	\Delta_{nm}(u,u) \equiv \frac{C_{nm}(u,u)-u^2}{u(1-u)}=\tUU_{nm}(u)+\tLL_{nm}(1-u)-1
\end{equation}
tends to the upper/lower tail dependence coefficients $\tUU_{nm}(1)$ and $\tLL_{nm}(1)$ when $u$ tends to $1$ resp. $0$.
Intuitively, the presence of $\Psi_N(u,v)$ in the covariance of $\overline{Y}$ above leads to a reduction of the 
number of effectively independent variables, 
but a more precise statement requires some further assumptions that we detail below. 

In the following, we will restrict to the case of {strong-}stationary random vectors, 
for which the copula $C_{nm}$ only depends on the lag $t=m-n${, i.e. $C_t\equiv C_{n,n+t}$}.   
The average of $\Delta_{nm}$ over $n,m$ can be turned into an average over $t$:
\begin{equation}\label{eq:avgDelta_t_homo}
	\Psi_N(u,v)=\sum_{t=1}^{N-1} (1-\frac{t}{N})\big(\Delta_{t}(u,v)+\Delta_{-t}(u,v)\big)
\end{equation}
with $\Delta_t(u,v) = \Delta_{n,n+t}(u,v)$. 
Note that in general $\Delta_t(u,v) \neq \Delta_{-t}(u,v)$, but clearly $\Delta_{t}(u,v)=\Delta_{-t}(v,u)$, which implies that
$\Psi_N(u,v)$ is symmetric in $u \leftrightarrow v$. 

We will assume in the following that the dependence encoded by $\Delta_t(u,v)$ has a limited range in time, 
or at least that it decreases sufficiently fast for the above sum to converge when $N \to \infty$. 
If the characteristic time scale for this dependence is $T$, we assume in effect that $T \ll N$. 
In the example worked out in Section~\ref{sec:example} below, one finds:
\[
\Delta_{t}(u,v) = f\left(\frac{t}{T}\right) \frac{A(u,v)}{I(u,v)}, \qquad I(u,v) \equiv \min(u,v)-uv
\]
where $f(\cdot)$ is a certain function. If $f(r)$ decays faster than $r^{-1}$, one finds (in the limit $T \gg 1$):
\[
\Psi_\infty(u,v) = \lim_{N \to \infty} \Psi_N(u,v) = T\, \frac{A(u,v)+A(v,u)}{I(u,v)} \int_0^\infty {\rm d}r f(r),
\]
with corrections at least of the order of $T/N$ when $N \gg T$.

\subsection{Limit properties}
We now define the process $\tilde{y}(u)$ as the limit of $\sqrt{N}\,\overline{Y}(u)$ when $N \to \infty$.
For a given $u$, it represents the asymptotics of the difference 
between the empirically determined cdf of the underlying $X$'s and the theoretical one, at the $u$-th quantile.
According to the Central Limit Theorem under weak dependences, 
it is Gaussian as long as the strong mixing coefficients, 
\[\fl
	\alpha_{SM}(t)=\sup_{\tau}\sup_{A,B}\left\{\big|\mathds{P}(A\cap B)-\mathds{P}(A)\mathds{P}(B)\big|: A\in \sigma(\{Z_n(u)\}_{n\leq \tau}),B\in \sigma(\{Z_n(v)\}_{n\geq \tau+t})\right\}
\]
associated to the sequence $\{Z_n(u)\}=\{Y_n(u)-u\}$, vanish at least as fast as $\mathcal{O}(t^{-5})$\footnote{This
condition means that the occurence of any two realizations of the underlying variable can be seen as independent for sufficiently long time between the realizations.
Since the copula induces a measure of probability on the Borel sets,
it amounts in essence to checking that $|C_t(u,v)-uv|$ converges quickly towards 0.
See Refs.~\cite{bradley2007introduction,chen2010nonlinearity,beare2010copulas} for definitions of 
$\alpha-$, $\beta-$, $\rho-$mixing coefficients and sufficient conditions on copulas for geometric mixing (fast exponential decay)
in the context of copula-based stationary Markov chains.}.
We will assume this condition to hold in the following. 
For example, this condition is met if the function $f(r)$ defined above decays exponentially, or if $f(r \geq 1)=0$.

The covariance of the process $\tilde{y}(u)$ is given by:
\begin{equation}\label{eq:Htheo}
	H(u,v)=\lim_{N\to\infty}N\,{\rm Cov}(\overline{Y}(u),\overline{Y}(v))=\left(\min(u,v)-uv\right)\left[1+\Psi_{\infty}(u,v)\right]
\end{equation}
and characterizes a Gaussian bridge since $\mathds{V}[\tilde{y}(0)]=\mathds{V}[\tilde{y}(1)]=0$, 
or equivalently $\mathds{P}[\tilde{y}(0)=y]=\mathds{P}[\tilde{y}(1)=y]=\delta(y)$.
Indeed, $I(u,v)=\min(u,v)-uv$ is the covariance function of the Brownian bridge, 
and $\Psi_{\infty}(u,v)$ is a non-constant scaling term.

By Mercer's theorem, the covariance $H(u,v)$ can be decomposed on its eigenvectors
and $\tilde{y}(u)$ can correspondingly be written as an infinite sum of Gaussian variables:
\begin{equation}\label{eq:y_sum_z}
	\tilde{y}(u)=\sum_{j=1}^{\infty}U_j(u)\sqrt{\lambda_j} \, z_j
\end{equation}
where $z_j$ are independent centered, unit-variance Gaussian variables, 
and the functions $U_j$ and the numbers $\lambda_j$ are solutions to the eigenvalue problem:
\begin{equation}
	\int_0^1 H(u,v) U_i(v)\,{\rm d}v=\lambda_i \, U_i(u)\quad\textrm{with}\quad\int_0^1 U_i(u)U_j(u)\,{\rm d}u= \delta_{ij}.
\end{equation}

In order to measure a limit distance between distributions, a norm over the space of continuous bridges needs to be chosen. 
Typical such norms are the norm-2 (sum of squares, as the bridge is always integrable), 
and the norm-sup (as the bridge always reaches an extremal value). Other norms, such as the Anderson-Darling test, can also 
be considered, see Ref.~\cite{darling1957kolmogorov}.

In practice, for every given problem, the covariance function in Equation~(\ref{eq:Htheo}) has a specific shape, 
since $\Psi_\infty(u,v)$ is copula-dependent. Therefore, contrarily to the case of independent random variables,
the GoF tests will not be characterized by universal (problem independent) distributions.

\subsection{Law of the norm-2 (Cram\' er-von-Mises)}
The norm-2 of the limit process is the integral of $\tilde{y}^2$ over the whole domain:
\numparts
\begin{equation}
	C\!M=\int_0^1\tilde{y}(u)^2\,{\rm d}u.
\end{equation}
In the representation (\ref{eq:y_sum_z}), it has a simple expression: 
\begin{equation}\label{eq:CM_sum_lambda}
	C\!M=\sum\limits_{j=1}^{\infty}\lambda_j z_j^2.
\end{equation}
\endnumparts
and its law is thus the law of an infinite sum of squared independent gaussian variables 
weighted by the eigenvalues of the covariance function. 
Diagonalizing $H$ is thus sufficient to find the distribution of $C\!M$, 
in the form of the Fourier transform of the characteristic function
\begin{equation}\label{eq:charctCM}
	\phi(t)=\mathds{E}\left[\e^{\mathrm{i}t\, C\!M}\right]
	       =\prod_{j}\left(1-2\mathrm{i}t\lambda_j\right)^{-\frac{1}{2}}.
\end{equation}
The hard task consists in finding the infinite spectrum of $H$ (or some approximations, if necessary).

Ordering the eigenvalues by decreasing amplitude, 
Equation~(\ref{eq:CM_sum_lambda}) makes explicit the decomposition of $C\!M$ over contributions of decreasing importance 
so that, at a wanted level of precision, only the most relevant terms can be retained.
In particular, if the top eigenvalue dominates all the others, we get the chi-square law with a single degree of freedom:
\begin{equation}
	\mathds{P}[C\!M\leq k]={\rm erf}\sqrt{\frac{k}{\lambda_0}}.
\end{equation}

Even if the spectrum cannot easily be determined but $H(u,v)$ is known, 
all the moments of the distribution can be computed exactly. For example:
\numparts\label{eq:momentsCvM}
\begin{eqnarray}
	\mathds{E}[C\!M]&=&{\rm Tr}{H} = \int_0^1H(u,u)\,{\rm d}u,\\
	\mathds{V}[C\!M]&\equiv& 2\,{\rm Tr}{H^2}=2\int\!\int_0^1H(u,v)^2\,{\rm d}u\,{\rm d}v.
\end{eqnarray}
\endnumparts

\subsection{Law of the supremum (Kolmogorov-Smirnov)}
The supremum of the difference between the empirical cdf of the sample and the target cdf under the null-hypothesis
has been used originally by Kolmogorov and Smirnov as the measure of distance.
The variable 
\begin{equation}
	K\!S=\sup_{u\in[0,1]}|\tilde{y}(u)|
\end{equation}
describes the limit behaviour of the GoF statistics. 
In the case where $1 + \Psi_\infty(u,v)$ can be factorized as $\sqrt{\psi(u)}\sqrt{\psi(v)}$, 
the procedure for obtaining the limiting distribution was worked out in \cite{anderson1952asymptotic}, 
and leads to a problem of a diffusive particle in an expanding cage, for which some results are known. 
There is however no general method to obtain the distribution of $K\!S$ for an arbitrary covariance function $H$. 

Nevertheless, if $H$ has a dominant mode, the relation (\ref{eq:y_sum_z}) becomes approximately: 
$
	\tilde{y}(u)=U_0(u)\sqrt{\lambda_0}z_0\equiv\kappa_0(u_0)z_0
$, and 
\begin{equation}
	K\!S=\sqrt{\lambda_0}|z_0|\sup_{u\in[0,1]}|U_0(u)|\equiv\kappa_0(u_0^*)|z_0|.
\end{equation}
The cumulative distribution function is then simply
\begin{equation}
	\mathds{P}[K\!S\leq k]={\rm erf}\left(\frac{k}{\sqrt{2}\kappa_0(u_0^*)}\right),\qquad k\geq 0.
\end{equation}
This approximation is however not expected to work for small values of $k$, since in this case $z_0$ must be small, 
and the subsequent modes are not negligible compared to the first one.
A perturbative correction --- working also for large $k$ only --- can be found when the second eigenvalue is small,
or more precisely when $\tilde{y}(u)=\kappa_0(u)z_0+\kappa(u)z_1$ with $\epsilon=\kappa/\kappa_0 \ll 1$.
The first thing to do is find the new supremum
\begin{equation}
	u^*=\arg\sup(\tilde{y}(u)^2)=u_0^*+\frac{\kappa'(u_0^*)}{|\kappa_0''(u_0^*)|}\frac{z_1}{z_0}.
\end{equation}
Notice that it is dependent upon $z_0,z_1$ so that $K\!S$ is no longer exactly the absolute value of a Gaussian.
However it can be shown (after lenghty but straightforward calculation) that, to second order in $\epsilon$, 
$\tilde{y}(u^*)$ \emph{remains  Gaussian}, albeit with a new width
\begin{equation}\label{eq:kappa_star}
	\kappa^* \approx \sqrt{\kappa_0^2+\kappa^2}>\kappa_0,
\end{equation}
where all the functions are evaluated at $u_0^*$. 
In fact, this approximation works also with more than two modes, provided
\begin{equation}
	\kappa(u)^2\equiv\sum_{j\neq 0}\lambda_jU_j(u)^2 \ll \kappa_0(u)^2=\lambda_0U_0(u)^2,
\end{equation}
in which case:
\begin{equation}
	\kappa^*\approx\sqrt{\sum_j\lambda_jU_j(u_0^*)}.
\end{equation}

\section{An explicit example: The log-normal volatility model}\label{sec:example}

In order to illustrate the above general formalism, 
we focus on the specific example of the product random variable $X=\sigma\xi$, 
with iid Gaussian residuals $\xi$ and log-normal stochastic standard-deviations $\sigma=\e^{\omega}$
(again we denote generically by $F$ the cdf of $X$).
Such models are common in finance to describe stock returns, as will be discussed in the next section.
For the time being, let us consider the case where the $\omega$'s are of zero mean, and covariance given by:
\begin{equation}\label{eq:modelLogNorm}
	{\rm Cov}(\omega_n \omega_{n+t}) = \Sigma^2 f\left(\frac{t}{T}\right), \qquad (t > 0).
\end{equation}

The pairwise copulas in the covariance of $\overline{Y}$ can be explicitely written in the limit of weak correlations, $\Sigma^2 \to 0$.
One finds:
\begin{eqnarray}
	C_t(u,v)-uv&=&\Sigma^2 f\left(\frac{t}{T}\right) \tilde{A}(u)\tilde{A}(v) + \mathcal{O}(\Sigma^4)\\\label{eq:def_A}
	\textrm{with}\quad \tilde{A}(u)&=&\int\limits_{-\infty}^{\infty}\varphi(\omega)\varphi'(\frac{F^{-1}(u)}{\e^{\omega}}){\rm d}\omega
\end{eqnarray}
where here and in the following $\varphi(\cdot)$ denotes the univariate Gaussian pdf, and $\Phi(\cdot)$ the Gaussian cdf.
The spectrum of $A(u,v)=\tilde{A}(u)\tilde{A}(v)$ consists in a single non-degenerate eigenvalue 
$\lambda^A={\rm Tr}A=\int_0^1\tilde{A}(u)^2\, {d}u$, 
and an infinitely degenerate null eigenspace. 
Assuming short-ranged memory, such that $f_\infty = \sum_{r=1}^\infty f(r) < +\infty$,
the covariance kernel reads:
\[
	H(u,v)=I(u,v)+ {2}T\Sigma^2 f_\infty \, A(u,v).
\]
Depending on the value of the parameters, the first term or the second term may be dominant. 
Note that one can be in the case of weak correlations ($\Sigma^2 \to 0$) but long range memory $T \gg 1$, 
such that the product $T\Sigma^2$ can be large (this is the case of financial time series, see below). 
If $T\Sigma^2$ is small, one can use perturbation theory around the Brownian bridge limit
(note that ${\rm Tr}I\approx 10 {\rm Tr}A$, see Appendix),
whereas if $T\Sigma^2$ is large, it is rather the Brownian term $I(u,v)$ that can be treated as a perturbation. 
Elements of the algebra necessary to set up these perturbation theories are given in the Appendix.

\begin{figure}
	\includegraphics[scale=0.42]{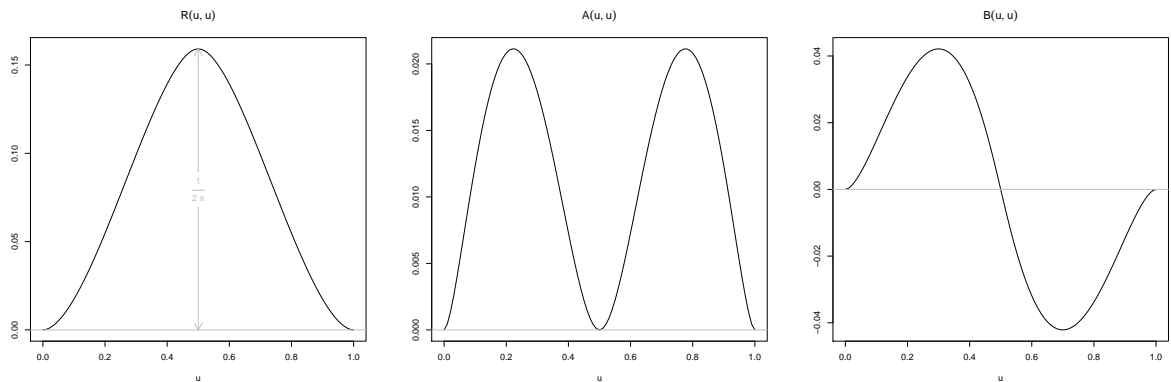}
	\caption{Copula diagonal of the log-normal volatility model: linear corrections to independence.
	         \textbf{Left:  } correction $R(u,u)$ due to correlation of the residuals (vertical axis in multiples of $\rho$)
	         \textbf{Middle:} correction $A(u,u)$ due to correlation of the log-vols (vertical axis in multiples of $\alpha$)
	         \textbf{Right: } correction $B(u,u)$ due to leverage effect (vertical axis in multiples of $-\beta$)}
	\label{fig:correctionsToIndep}
\end{figure}

It is interesting to generalize the above model to account for weak dependence between the residuals $\xi$ and between the residual and the volatility,
without spoiling the log-normal structure of the model. We therefore write:
\[
	X_0=\xi_0\e^{\omega_0};\quad X_t=\xi_t\e^{\alpha_t\omega_0+\beta_t\xi_0+\sqrt{1-\alpha_t^2-\beta_t^2}\omega_t}
	\quad\textrm{with}\quad
	\mathds{E}[\xi_0\xi_t]=r_t
\]
where all the variables are $\mathcal{N}(0,1)$, so that in particular 
\begin{eqnarray*}
	\rho_t=&{\rm Corr}(X_0,X_t)=r_t(1+\beta_t^2)\e^{\alpha_t-1}\\
	       &{\rm Corr}(X_0^2,X_t^2)=\frac{\left(1+2r_t^2(1+10\beta_t^2+8\beta_t^4)+4\beta_t^2\right)\e^{4\alpha_t}-1}{3\e^4-1}\\
	       &{\rm Corr}(X_0,X_t^2)=2\beta_t\frac{\left(1+2r_t^2(1+2\beta_t^2)\right)\e^{2\alpha_t-\frac{1}{2}}}{\sqrt{\e^4-1}}
\end{eqnarray*}
The univariate marginal distributions of $X_0$ and $X_t$ are identical and their cdf is given by the integral
\begin{equation}
	F(x)=\int_{-\infty}^{\infty}\varphi(\omega)\Phi(\frac{x}{\e^{\omega}}) {\rm d}\omega.
\end{equation}
Expanding the bivariate cdf (or the copula) in the small dependence parameters $\alpha_t,\beta_t,\rho_t$ around $(0,0,0)$, we get
\begin{eqnarray}\label{eq:cop_perturb1}
	C_{t}(u,v)-uv  &\approx& \alpha_t A(u,v) - \beta_t B(u,v) + \rho_t R(u,v)\\\nonumber\label{eq:cop_perturb2}
	               &\approx& \alpha_t\tilde{A}(u)\tilde{A}(v)-\beta_t\tilde{R}(u)\tilde{A}(v)+\rho_t\tilde{R}(u)\tilde{R}(v)
\end{eqnarray}
where $\tilde{A}(u)$ was defined above in Equation~(\ref{eq:def_A}), and 
\begin{eqnarray}
	\tilde{R}(u)&=&\int_{-\infty}^{\infty}\varphi(\omega)\varphi(\frac{F^{-1}(u)}{\e^\omega}){\rm d}\omega =\tilde{R}(1-u).
\end{eqnarray}
The contributions of $A(u,v), B(u,v)$ and $R(u,v)$ on the diagonal are illustrated in Figure~\ref{fig:correctionsToIndep}.
Notice that the term $B(u,v)$ (coming from cross-correlations between $\xi_0$ and $\omega_t$, i.e. the so-called leverage effect, see below) breaks the symmetry $C_{t}(u,v) \neq C_{t}(v,u)$.

We now turn to a numerical illustration of our theory, in the simple case where only volatility correlations are present 
(i.e. $\beta_t =\rho_t =0$ in Equation~(\ref{eq:cop_perturb1}) above).
We furthermore assume a Markovian dynamics for the log-volatilities:
\begin{equation}\label{eq:markov_logvol}
X_n=\xi_n\e^{\omega_{n}-\mathds{V}[\omega]}, \quad\textrm{with}\quad\omega_{n+1} = g \omega_n + \Sigma \eta_n,
\end{equation}
where $g < 1$ and $\eta_n$ are iid Gaussian variables of zero mean and unit variance. In this case, 
\begin{equation}
	\alpha_t=\,\,{\rm Cov}(\omega_n \omega_{n+t}) = \frac{\Sigma^2}{1-g^2}\,g^t.
\end{equation}
In the limit where $\Sigma^2 \ll 1$, the weak dependence expansion holds and one finds explicitly:
\begin{equation}\label{eq:weak_dep_kernel}
	H(u,v)=I(u,v)+ {2}\frac{g \Sigma^2}{(1-g)^2(1+g)} \, A(u,v).
\end{equation}
In order to find the limit distribution of the test statistics, we procede by Monte-Carlo simulations.
The range $[0,1]^2$ of the copula is discretized on a regular lattice of size $(M\times M)$.
The limit process is described as a vector with $M$ components and built from Equation~(\ref{eq:y_sum_z}) as  
$\mathbf{\tilde{y}}=U\Lambda^{\frac{1}{2}}\mathbf{z}$ 
where the diagonal elements of $\Lambda$ are the eigenvalues of $H$ (in decreasing order),
and the columns of $U$ are the corresponding eigenvectors. 
Clearly, ${\rm Cov}(\mathbf{\tilde{y}},\mathbf{\tilde{y}})=U\Lambda U^{\dagger}=H$.

\begin{figure}
	\begin{indented}
	\item[]
	\includegraphics[scale=0.45]{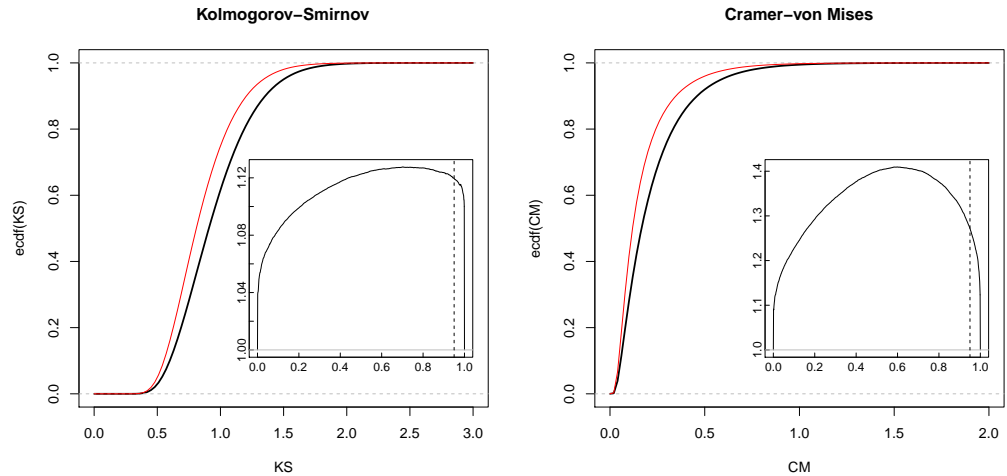}
	\end{indented}
	\caption{Markovian model -- \textbf{Left:} Cumulative distribution function of the supremum of $\tilde{y}(u)$.
	         \textbf{Right:} Cumulative distribution of the norm-2 of $\tilde{y}(u)$.
	         The cases of independent drawings (thin red) and dependent drawings (bold black) are compared. 
	         The dependent observations are drawn according to the weak-dependence kernel (\ref{eq:weak_dep_kernel}) with parameters $g=0.88, \Sigma^2=0.05$. 
	         \textbf{Insets:} The effective reduction ratio $\sqrt{\frac{N}{N_{\mbox{\scriptsize{eff}}}(u)}}=\frac{\textrm{ecdf}^{-1}(u)}{\textrm{cdf}_{\mbox{\scriptsize{L}}}^{-1}(u)}$ 
	                          where $\textrm{L}=\textrm{KS,CM}$.
	                          The dashed vertical line is located at the 95-th centile and thus indicates the reduction ratio corresponding to the p-value $p=0.05$ (as the test is unilateral).}
	\label{fig:ecdf_sim}
\end{figure}

For each Monte-Carlo trial, $M$ independent random values are drawn 
from a standard Gaussian distribution and collected in $\mathbf{z}$.
Then $\mathbf{y}$ is computed using the above representation. This allows one to determine the two relevant statistics:
\begin{eqnarray*}
	K\!S&=&\max_{u=1\ldots M}|\tilde{y}_u|\\
	C\!M&=&\frac{1}{M}\sum_{u=1}^M\tilde{y}_u^2=\frac{1}{M}\mathbf{\tilde{y}}^{\dagger}\mathbf{\tilde{y}}=\frac{1}{M}\mathbf{z}^{\dagger}\Lambda\mathbf{z}.
\end{eqnarray*}
The empirical cumulative distribution functions of the statistics for a large number of trials are shown in Figure~\ref{fig:ecdf_sim} 
together with the usual Kolmogorov-Smirnov and Cram\' er-von-Mises limit distributions 
corresponding to the case of independent variables.

\begin{figure}
	\begin{indented}
	\item[]
	\includegraphics[scale=0.45]{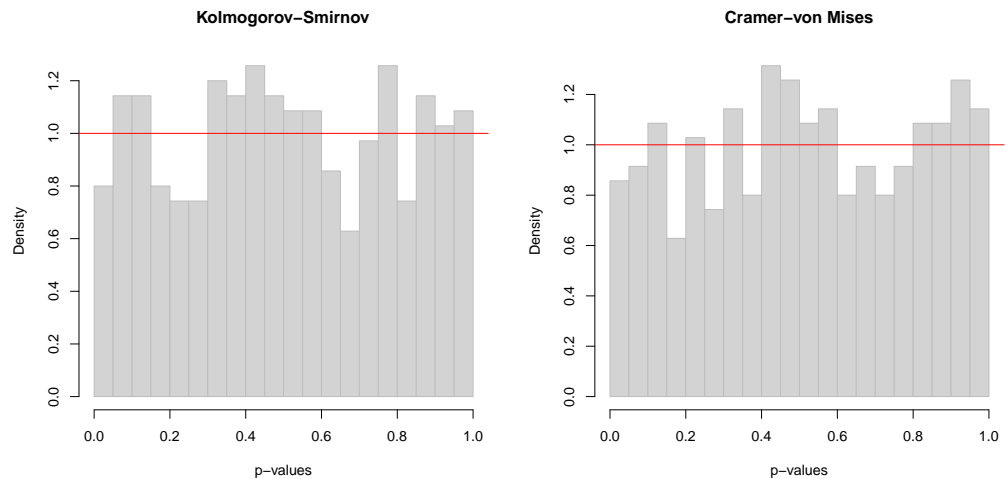}
	\end{indented}
	\caption{Histogram of the p-values in the GoF test on simulated data,
	according to Equation~(\ref{eq:markov_logvol}).
	Uniform distribution of the p-values of a test indicates that the correct law of the statistics is used.}
	\label{fig:KS_pvals_sim}
\end{figure}

In order to check the accuracy of the obtained limit distribution, 
we generate $350$ series of $N=2500$ dates according to Equation~(\ref{eq:markov_logvol}).
For each such series, we perform two GoF tests, namely KS and CM, and calculate the corresponding p-values. 
By construction, the p-values of a test should be uniformly distributed 
if the underlying distribution of the simulated data is indeed the same as the hypothesized distribution.
In our case, when using the usual KS and CM distributions for independent data,
the p-values are much too small and their histogram is statistically not compatible with the uniform distribution.
Instead, when using the appropriate limit distribution found by Monte-Carlo and corresponding to the correlation kernel (\ref{eq:weak_dep_kernel}),
the calculated p-values are uniformly distributed, as can be visually seen on Figure~\ref{fig:KS_pvals_sim}, 
and as revealed statistically by a KS test (on the KS test!), comparing the 350 p-values to ${\rm H}_0: p\sim\mathcal{U}[0,1]$.

\begin{figure}[b]
	\includegraphics[scale=0.35,trim=425 0 0 0,clip]{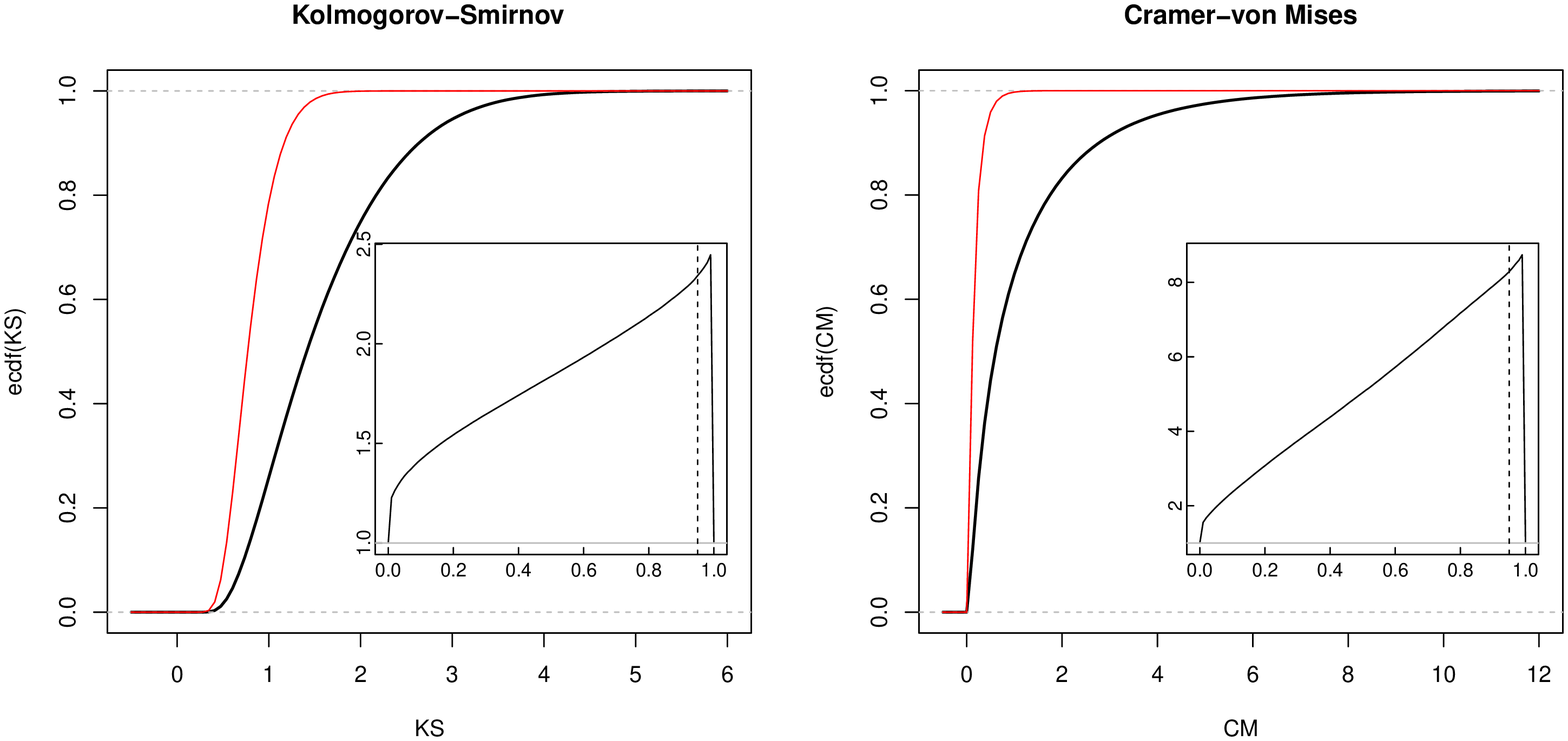}
	\includegraphics[scale=0.53,trim=280 0 0 325,clip]{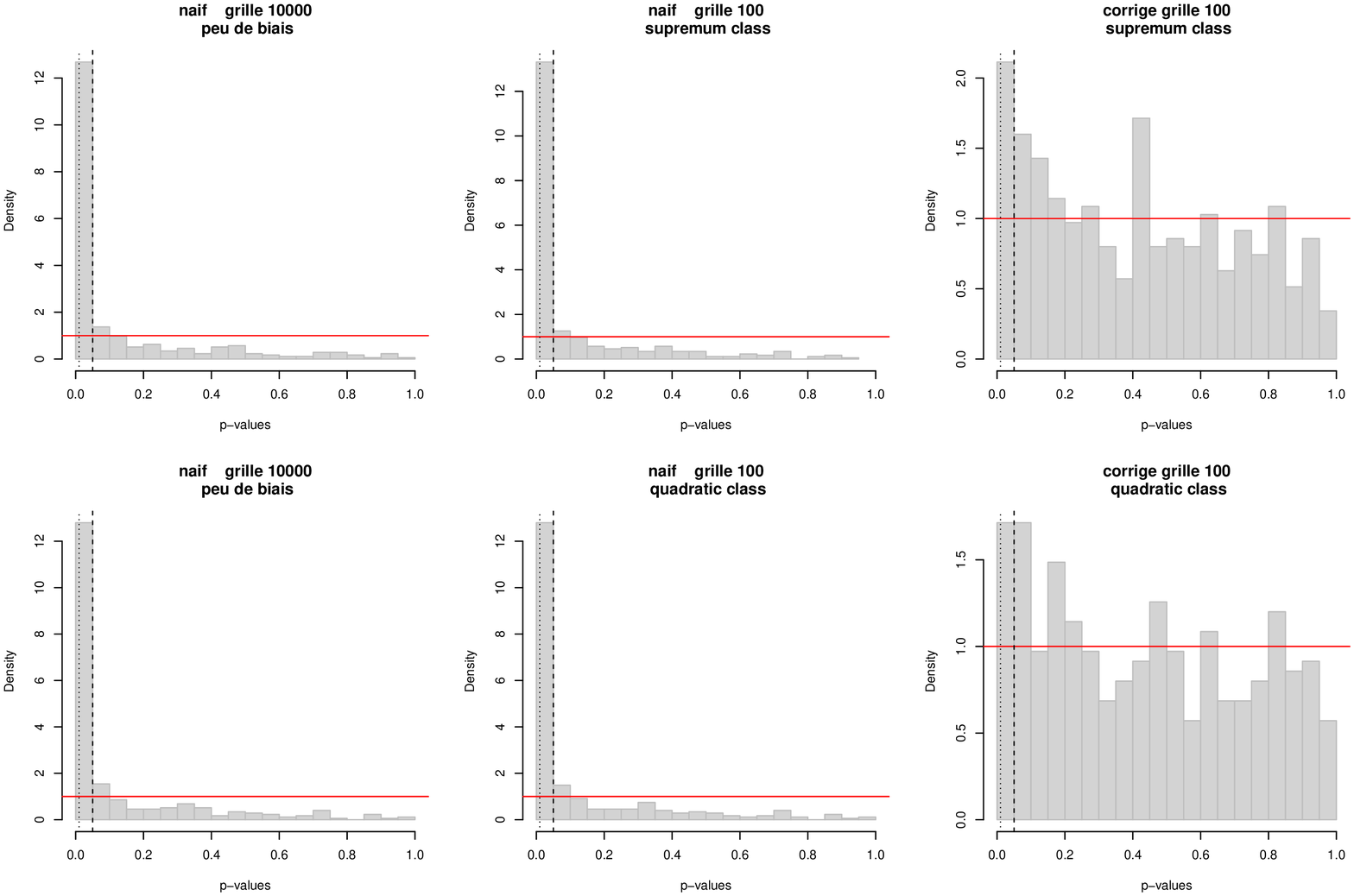}
	\caption{Fractional Brownian Motion -- \textbf{Left:} Cumulative distribution function of the norm-2 of $\tilde{y}(u)$, see Fig.~\ref{fig:ecdf_sim} for full caption.
	         \textbf{Middle-Right:} Histogram of the p-values in the CM test on simulated data, using the iid CM distribution (middle) and the 
	         prediction obtained assuming short range correlations (right). 
	         The dependent observations are drawn according to (\ref{eq:alpha_FBM}) with parameters $\nu=\frac{2}{5}, \Sigma^2=1, N = 1500$. }
	\label{fig:ecdf_FGN}
\end{figure}

\paragraph{}
If, instead of the AR(1) (Markovian) prescription (\ref{eq:markov_logvol}), the dynamics of the $\omega_n$ is given by
a Fractional Gaussian Noise (i.e. the increments of a fractional Brownian motion) \cite{mandelbrot1968fractional}
with Hurst index $\frac{2-\nu}{2}>\frac{1}{2}$, the log-volatility has a long ranged autocovariance
\begin{equation}\label{eq:alpha_FBM}
	\alpha_t={\rm Cov}(\omega_n,\omega_{n+t})=\frac{\Sigma^2}{2}\left((t+1)^{2-\nu}-2t^{2-\nu}+|t-1|^{2-\nu}\right),\quad t\geq 0
\end{equation}
that decays as a power law $\propto (2-3\nu+\nu^2)t^{-\nu}$ as $t\to\infty$, corresponding to long-memory, 
therefore violating the hypothesis under which the above theory is correct.
Still, we want to illustrate that the above methodology leads to a meaningful improvement of the test, even in the case of long-ranged dependences.
The corresponding covariance kernel of the $X$s, 
\begin{equation}\label{eq:kernel_FBM}
	H(u,v)=I(u,v)+2\Sigma^2A(u,v)\sum_{t=1}^{N}\left(1-\frac{t}{N}\right)\alpha_t,
\end{equation}
is used in a Monte-Carlo simulation like in the previous case in order to find the appropriate 
distribution of the test statistics $KS$ and $CM$ (shown in Figure~\ref{fig:ecdf_FGN}, see caption for the choice of parameters).
We again apply the GoF tests to simulated series, and compute the p-values according to the theory above.
As stated above, our theory is designed for short-ranged dependences whereas the FGN process is long-ranged. The p-values are therefore 
not expected to be uniformly distributed. Nevertheless, the distribution of the p-values is significantly corrected toward the uniform distribution, see Figure~\ref{fig:ecdf_FGN}:
with the naive CM distribution (middle), the obtained p-values are localized around zero, suggesting that the test is strongly negative. 
If instead we use our prediction for short-range dependences (right), we find a clear improvement, as the
p-values are more widely spread on $[0,1]$ (but still not uniformly).

\newpage
\section{Application to financial time series}\label{sec:application}

\subsection{Stylized facts of daily stock returns}

One of the contexts where long-ranged persistence is present is time series of financial asset returns. 
At the same time, the empirical determination of the distribution of these returns is of utmost importance, 
in particular for risk control and derivative pricing. 
As we will see, the volatility correlations are so long-ranged that the number of effectively independent observations is strongly reduced, 
in such a way that the GoF tests are not very tight, 
even with time series containing thousands of raw observations. 

It is well-known that stock returns exhibit dependences of different kinds:
\begin{itemize}
\item at relatively high frequencies (up to a few days), returns show weak, but significant negative linear auto-correlations (see e.g. \cite{avramov2006liquidity});
\item the absolute returns show persistence over very long periods, an effect called multiscale volatility clustering and
for which several interesting models have been proposed in the last ten years \cite{pasquini1999multiscaling,calvetfisher,bacrymuzy,zumbach2001heterogeneous,lynch2003market,borland2005dynamics};
\item past negative returns favor increased future volatility , an effect that goes under the name of ``leverage correlations'' in the
literature \cite{bouchaud2001more,perello2004multiple,pochart2002skewed,eisler2004multifractal,ahlgren2007frustration,reigneron2011principal}.
\end{itemize}

Our aim here is neither to investigate the origin of these effects 
and their possible explanations in terms of behavioral economics, nor to propose a new 
family of models to describe them. We rather want to propose a new way to characterize and 
measure the full structure of the temporal dependences of returns based on copulas, and
extract from this knowledge the quantities needed to apply  GoF tests to financial times series.

Throughout this section, the empirical results are based on a data set consisting of
the daily returns of the stock price of listed large cap US companies.
More precisely we keep only the 376 names present in the S\&P-500 index constantly over the five years period 2000--2004, corresponding to $N=1256$ days.
The individual series are standardized, but this does not change the determination of copulas, 
that are invariant under increasing and continuous transformations of the marginals.

\subsection{Empirical self-copulas}

For each $(u,v)$ on a lattice, we determine the lag dependent ``self-copula'' $C_t(u,v)$ by assuming stationarity, 
i.e. that the pairwise copula $C_{nm}(u,v)$ only depends on the time lag $t=m-n$. 
We also assume that all stocks are characterized by the same self-copula, 
and therefore an average over all the stock names in the universe is done in order to remove noise. 
Both these assumptions are questionable and we give at the end of this section an insight on
how non-stationarities as well as systematic effects of market cap, liquidity, tick size, etc, 
can be accounted for.

\begin{figure}
	\begin{indented}
	\item[]
	\includegraphics[scale=0.5,trim=0    0 1620 0,clip]{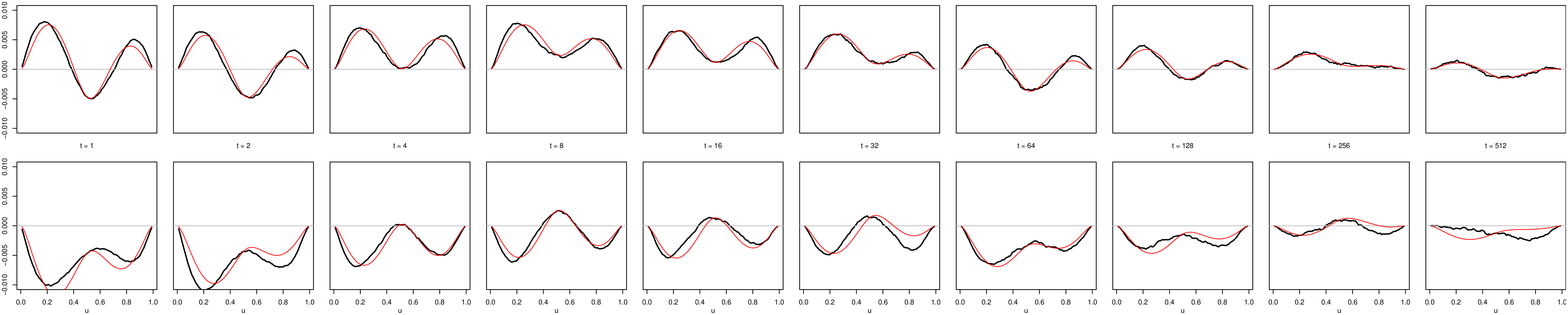}
	\includegraphics[scale=0.5,trim=550  0 1075 0,clip]{fig4.eps}
	\includegraphics[scale=0.5,trim=915 0  720  0,clip]{fig4.eps}
	\includegraphics[scale=0.5,trim=1450 0  175  0,clip]{fig4.eps}
	\end{indented}
	\caption{Diagonal (top) and anti-diagonal (bottom) of the self-copula for different lags;
	the product copula has been subtracted. A fit with Equation~(\ref{eq:cop_perturb1}) is shown in thin red. Note that the $y$ scale is small, confirming that 
	the weak dependence expansion is justified. The dependence is still significant even for $t \sim 500$ days!}
	\label{fig:diag-adiag-tau}
\end{figure}

The self-copulas are estimated non-parametrically with a bias correction\footnote{Details on the copula estimator and the bias issue are given in appendix.}, 
then fitted to the parametric family of log-normal copulas introduced in the previous section. 
We assume (and check a posteriori) that the weak dependence expansion holds, leaving us with three functions of time, 
$\alpha_t$, $\beta_t$ and $\rho_t$, to be determined. 
We fit for each $t$ the copula diagonal $C_t(u,u)$ to Equation~(\ref{eq:cop_perturb1}) above, 
determine  $\alpha_t$, $\beta_t$ and $\rho_t$, and test for consistency on the anti-diagonal $C_t(u,1-u)$. 
Alternatively, we could determine these coefficients to best fit $C_t(u,v)$ in the whole $(u,v)$ plane, 
but the final results are not very different. 
The results are shown in Figure~\ref{fig:diag-adiag-tau} for lags $t=1,8,32,256$ days. 
Fits of similar quality are obtained up to $t=512$.  

Before discussing the time dependence of the fitted coefficients  $\alpha_t$, $\beta_t$ and $\rho_t$, 
let us describe how the different effects show up in the plots of the diagonal and anti-diagonal copulas. 
The contribution of the linear auto-correlation can be directly observed at the central point $C_t(\frac{1}{2},\frac{1}{2})$ of the copula.
It is indeed known \cite{chicheportiche2010joint} that for any pseudo-elliptical model (including the present log-normal framework) one has:
\[
	C_t(\case{1}{2},\case{1}{2})=\frac{1}{4}+\frac{1}{2\pi}\arcsin \rho_t.
\]
Note that this relation holds beyond the weak dependence regime. 
If $\betaB_t=C_t(\frac{1}{2},\frac{1}{2})-\frac{1}{4}$ 
--- this is in fact Blomqvist's beta coefficient \cite{blomqvist1950measure} --- 
the auto-correlation is measured by $\rho_t=\sin(2\pi \betaB_t)$.

The volatility clustering effect can be visualized in terms of the diagonals of the self-copula;
indeed, the excess (unconditional) probability of large events following previous large events of the same sign is 
$\big(C_t(u,u)-u^2\big)$ with $u<\frac{1}{2}$ for negative returns, and $u>\frac{1}{2}$ for positive ones.
On the anti-diagonal, the excess (unconditional) probability of large positive events following large negative ones is, 
for small $u<\frac{1}{2}$, the upper-left volume
$\big(C_t(u,1)-u\cdot 1\big)-\big(C_t(u,1\!-\!u)-u(1\!-\!u)\big)=u(1\!-\!u)-C_t(u,1\!-\!u)$ 
and similarly the excess probability of large negative events following large positive ones is 
the same expression for large $u>\frac{1}{2}$ (lower-right volume).
As illustrated on Figure~\ref{fig:diag-adiag-tau}, these four quadrants exceed the independent case prediction, 
suggesting a genuine clustering of the amplitudes, measured by $\alpha_t$.
Finally, an asymmetry is clearly present: 
the effect of large negative events on future amplitudes is stronger than the effect of previous positive events.
This is an evidence for the leverage effect: negative returns cause a large volatility, 
which in turn makes future events (positive or negative) to be likely larger. This effect is captured by the coefficient $\beta_t$.

\begin{figure}
	\begin{indented}
	\item[]
	\includegraphics[scale=0.4]{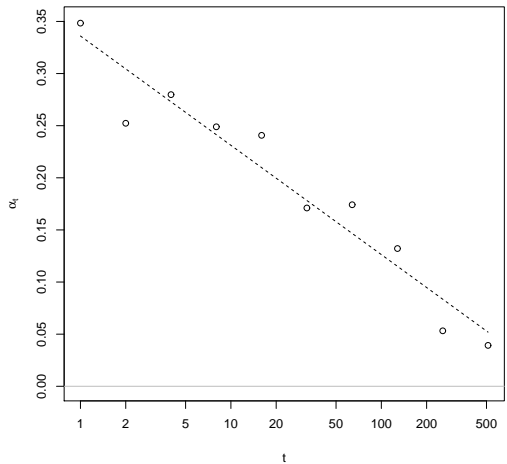}
	\end{indented}
	\caption{Auto-correlation of the volatilities, for lags  ranging from $1$ to $768$ days. 
		Each point represents the value of $\alpha_t$ extracted from a fit of the empirical copula diagonal 
		at a given lag to the relation (\ref{eq:cop_perturb1}). We also show the fit to a multifractal model, 
		$\alpha_t=-\Sigma^2\log\frac{t}{T}$, with $\Sigma^2=0.046$ and $T=1467$ days.}
	\label{fig:alpha_t}
\end{figure}

The evolution of the coefficients  $\alpha_t$, $\beta_t$ and $\rho_t$ for different lags reveals the following properties:
i) the linear auto-correlation $\rho_t$ is short-ranged (a few days), and negative;
ii) the leverage parameter $\beta_t$ is short-ranged and, as is well known, negative, revealing the asymmetry discussed above;
iii) the correlation of volatility is {\it long-ranged} and of relatively large positive amplitude (see Figure~\ref{fig:alpha_t}), in line with the known long range 
volatility clustering. 
More quantitatively, we find that the parameter $\alpha_t$ for lags ranging from $1$ to $768$ days
is consistent with an effective relation 
well fitted by the ``multifractal''  \cite{muzy2001multifractal,calvetfisher,lux,bacrymuzy} prediction for the 
volatility autocorrelations: $\alpha_t=-\Sigma^2\log\frac{t}{T}$, with an amplitude $\Sigma^2=0.046$ and a horizon $T=1467$ days 
consistent, in order of magnitude, with previous determinations.

\begin{figure}
	\begin{indented}
	\item[]
	\includegraphics[scale=0.5]{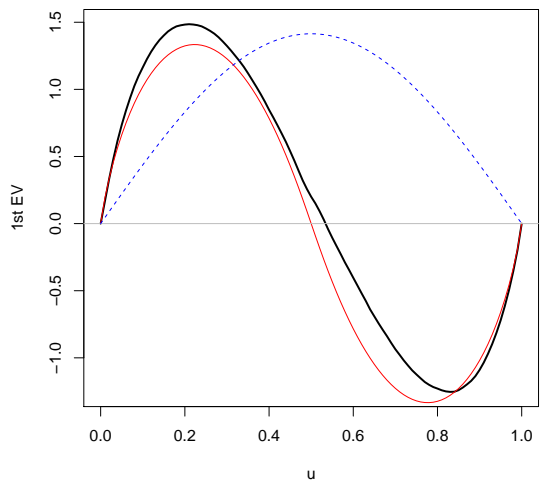}
	\end{indented}
	\caption{\textbf{Bold black:}
		 The first eigenvector of the empirical kernel $H(u,v)=I(u,v) \left[1 +\Psi_N(u,v)\right]$.
	         \textbf{Plain red:}
	         The function $\tilde{A}(u)$ (normalized), corresponding to the pure effect of volatility clustering in a log-normal model, in the
	         limit where the Brownian bridge contribution $I(u,v)$ becomes negligible. 
	         \textbf{Dashed blue:}
	         The largest eigenmode $\ket{1}=\sqrt{2}\sin(\pi u)$ of the independent kernel $I(u,v)$.
	         }
	\label{fig:eigen_vects}
\end{figure}

The remarkable point, already noticed in previous empirical works on multifractals \cite{muzy2001multifractal,duchon2010forecasting}, 
is that the horizon $T$, beyond which the volatility correlations vanish, is found to be extremely long. 
In fact, the extrapolated horizon $T$ is larger than the number of points of our sample $N$! 
This long correlation time has two consequences: 
first, the parameter ${2}T\Sigma^2 f_\infty$ that appears in the kernel $H(u,v)$ is large, $\approx {135}$. 
This means that the dependence part $T\Sigma^2 f_\infty \, A(u,v)$ is dominant over the independent Brownian bridge part $I(u,v)$. 
This is illustrated in Figure~\ref{fig:eigen_vects}, where we show the first eigenvector of $H(u,v)$, 
which we compare to the non-zero eigenmode of $A(u,v)$, and to the first eigenvector of $I(u,v)$. Second, the hypothesis of a stationary process, which 
requires that $N \ll T$, is not met here, so we expect important preasymptotic corrections to the above theoretical results.

\subsection{Monte-Carlo estimation of the limit distributions}

Since $H(u,v)$ is copula-dependent, and considering the poor analytical progress made 
about the limit distributions of $K\!S$ and $C\!M$ in cases other than independence, 
the asymptotic laws will be computed numerically by Monte-Carlo simulations (like in the example of Section~\ref{sec:example})
with the empirically determined $H(u,v)$. 

\begin{figure}
	\begin{indented}
	\item[]
	\includegraphics[scale=0.45]{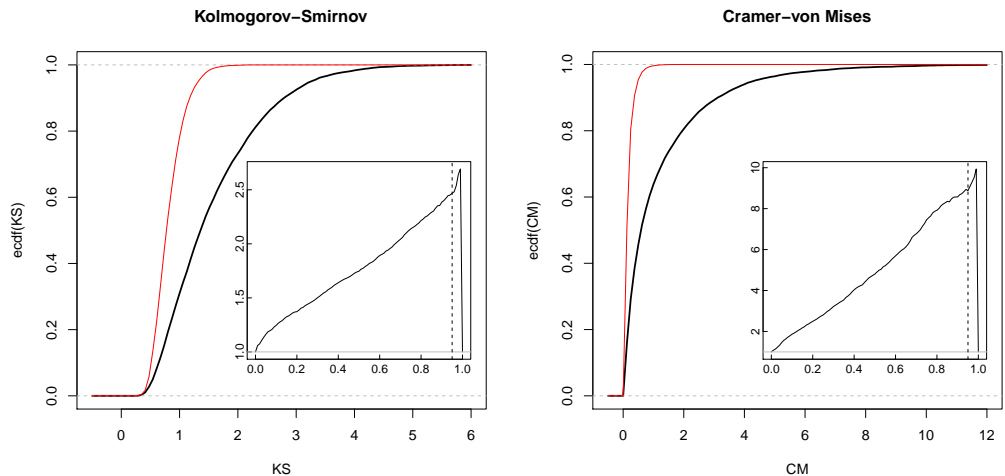}
	\end{indented}
	\caption{\textbf{Left:} Cumulative distribution function of the supremum of $\tilde{y}(u)$.
	         \textbf{Right:} Cumulative distribution of the norm-2 of $\tilde{y}(u)$.
	         The cases of independent drawings (thin red) and dependent drawings (bold black) are compared. 
	         The dependent observations are drawn according to the empirical average self-copula of US stock returns in 2000-2004. 
	         \textbf{Insets:} The effective reduction ratio $\sqrt{\frac{N}{N_{\mbox{\scriptsize{eff}}}(u)}}=\frac{\textrm{ecdf}^{-1}(u)}{\textrm{cdf}_{\mbox{\scriptsize{L}}}^{-1}(u)}$ 
	                          where $\textrm{L}=\textrm{KS,CM}$.}
	\label{fig:ecdf_stats}
\end{figure}

The empirical cumulative distribution functions of the statistics for a large number of trials are shown in Figure~\ref{fig:ecdf_stats} 
together with the usual Kolmogorov-Smirnov and Cram\' er-von-Mises limit distributions 
corresponding to the case of independent variables.
One sees that the statistics adapted to account for dependences are stretched to the right,
meaning that they accept higher values of $K\!S$ or $C\!M$ (i.e. measures of the difference between the true and the empirical distributions).
In other words, the outcome of a test based on the usual $K\!S$ or $C\!M$ distributions is much more likely to be negative, 
as it will consider ``high'' values (like 2--3) as extremely improbable, whereas a test that accounts for the strong dependence in the time series 
would still acccept the null-hypothesis for such values.

As an illustration, we apply the test of Cram\'er-von Mises to our dataset, comparing the empirical univariate distributions of stock returns 
to a simple model of log-normal stochastic volatility
\begin{equation}\label{eq:lognorm_vol}
	X=\e^{s\omega-s^2}\xi\qquad\textrm{where}\quad\xi,\omega\stackrel{\textrm{\tiny iid}}{\sim}\mathcal{N}(0,1).
\end{equation}
The volatility of volatility parameter $s$ can be calibrated from the time series $\{x_t\}_t$ as
\begin{equation}\label{eq:volvol}
	s^2=\log\left(\frac{2}{\pi}\frac{\vev{x_t^2}_t}{\vev{x_t}_t^2}\right).
\end{equation}
We want to test the hypothesis that the log-normal model with a unique value of $s$ {\it for all stocks} is compatible with the data. 
In practice, for each stock $i$, $s_i$ is chosen as the average of (\ref{eq:volvol}) over all \emph{other} stocks
in order to avoid endogeneity issues and biases in the calculations of the p-values of the test. 
$s_i$ is found to be $\approx 0.5$ and indeed almost identical for all stocks. 
Then the GoF statistic $C\!M$ is computed for each stock $i$ and the corresponding p-value is calculated. 

\begin{figure}
	\begin{indented}
	\item[]
	\includegraphics[scale=0.5,trim=350 0 0 400,clip]{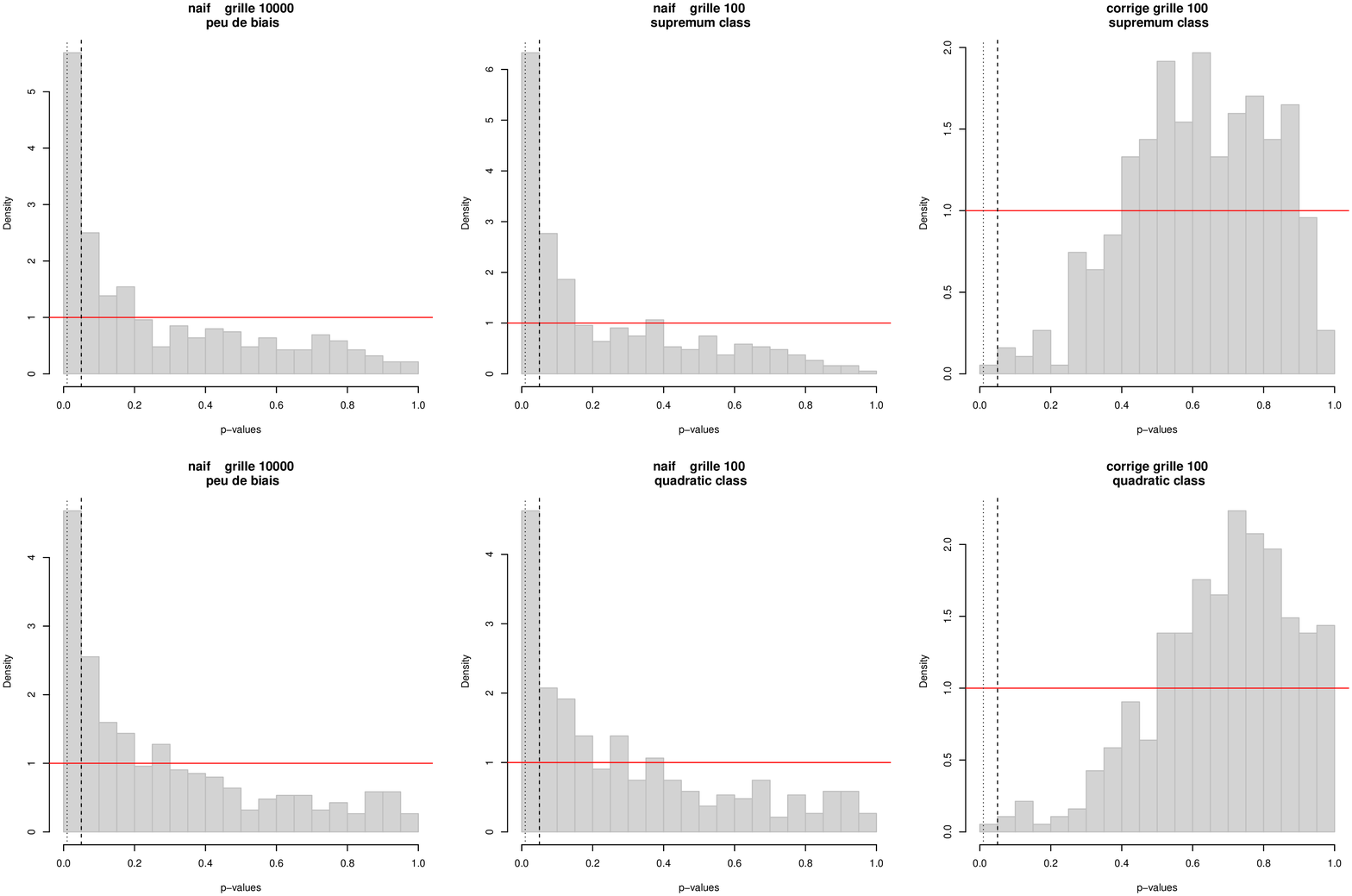}
	\end{indented}
	\caption{Histogram of the p-values in a Cram\'er-von Mises-like test,
	         see the text for the test design.
	         \textbf{Left: } when using the standard Cram\'er-von Mises law, the obtained p-values are
	         far from uniformly distributed and strongly localized under the threshold $p=0.05$ 
	         (dashed vertical line) occasioning numerous spurious rejections.
	         \textbf{Right: } when using the modified law taking dependences into account, 
	         the test rejects the hypothesis of an identical distribution for all stocks much less often.
	         }
	\label{fig:KS_pvals}
\end{figure}

Figure~\ref{fig:KS_pvals} shows the distribution of the p-values, as obtained by using the usual asymptotic Cram\'er-von Mises distribution 
for independent samples (left) and the modified version allowing for dependence (right).
We clearly observe that the standard Cram\'er-von Mises test strongly rejects the hypothesis of a common log-normal model, 
as the corresponding p-values are concentrated around zero, which leads to an excessively high rejection rate. 
The use of the generalized Cram\'er-von Mises test for dependent variables greatly improves the situation, with in fact now too many high values of $p$. 
Therefore, {\it the hypothesis of a common log-normal model for all stocks cannot be rejected} when the long-memory of volatility is taken into account. 
The overabundant large values of $p$ may be due to the fact that all stocks are in fact exposed to a common volatility factor 
(the ``market mode''), which makes the estimation of $s$ somewhat endogeneous and generates an important bias. 
Another reason is that the hypothesis that the size of the sample $N$ is much larger than the correlation time $T$ does not hold for our sample, 
and corrections to our theoretical results are expected in that case.\footnote{Note that in practice,
we have estimated $\Psi_N(u,v)$ by summing the empirically determined copulas up to $t_{\max} = 512$, 
which clearly underestimates the contribution of large lags.}
It would be actually quite interesting to extend the above formalism to the long-memory case, where $T \gg N \gg 1$.

\subsection{{Beyond stationarity and universality}}

The assumption that financial time series are stationary is far from granted.
In fact, several observations suggest that financial markets operate at the frontier between stationarity and non-stationarity. The multifractal model, 
for example, assumes that correlations are decaying marginally slowly (as a logarithm), which technically corresponds to the boundary between 
stationary models (when decay is faster) and non-stationary models. Furthermore, as stated above, the horizon $T$ that appears in the model is 
empirically found to be very long (on the order of several years) and therefore very difficult to measure precisely. Other models, like the ``multi-scale'' GARCH \cite{zumbach2001heterogeneous,borland2005multi}, 
point in the same direction: when these models are calibrated on financial data, the parameters are always found to be very close to the limit of 
stability of the model (see \cite{borland2005multi} for a discussion of this point). 

Furthermore, what is relevant in the present context is strong stationarity, i.e. the stationarity of the full self-copula. Hence, testing for 
stationarity amounts to comparing copulas, which amounts to a GoF for two-dimensional variables \ldots for which general statistical tools are missing, 
even in the absence of strong dependence! So in a sense this problem is like a snake chasing its own tail. 
A simple (and weak) argument is to perform the above study in different periods. When fitting the multifractal model to the self-copulas, we find the 
following values for $(\Sigma^2,\ln T)$: $(0.032,6.39)$ in 1995-1999, $(0.046,7.29)$ in 2000-2004, $(0.066,8.12)$ in 2000-2009 and $(0.078,7.86)$ in 2005-2009.
These numbers vary quite a bit, but this is exactly what is expected with the multifractal model itself when the size of the time series $N$ is not much larger than the
horizon $T$! (see \cite{muzy2006extreme,bacry2006asset} for a detailed discussion of the finite $N$ properties of the model). 

As of the universality of the self-copula across the stocks, it is indeed a reasonable assumption that allows one to average over the stock ensemble.
We compared the averaged self-copula on two subsets of stocks obtained by randomly dividing the ensemble, and found the resulting copulas to be hardly distinguishable.
This can be understood if we consider that all stocks are, to a first approximation, driven by a common force --- the ``market'' --- and subject to idiosyncratic fluctuations.
We know that this picture is oversimplified, and that systematic effects of sector, market cap, etc. are expected and could in fact be treated separately 
inside the framework that we propose. Such issues are related to the cross-sectional non-linear dependence structure of stocks, and are far beyond the 
objectives of this article.

\section{Conclusion}
The objectives of this paper were twofold: on the theoretical side, we introduced a framework for the study of statistical tests of Goodness-of-Fit 
with dependent samples; on the empirical side, we presented new measurements as well as phenomenological models for 
non-linear dependences in time series of daily stock returns. Both parts heavily rely on the notion of bivariate self-copulas. 

In summary, GoF testing on persistent series cannot be universal as is the case for iid variables, but requires a careful estimation of the self-copula at all lags.
Correct asymptotic laws for the Kolmogorov-Smirnov and Cram\'er-von Mises statistics can be found as long as dependences are short ranged, i.e. $T\ll N$.
From the empirical estimation of the self-copula of US stock returns, long-ranged volatility clustering with multifractal properties is observed
as the dominant contribution to self-dependence, in line with previous studies. 
However, subdominant modes are present as well and a precise understanding of those involves an in-depth study of the spectral properties of the 
correlation kernel $H$.

One of the remarkable consequences of the long-memory nature of the volatility is that the number of effectively independent observations is significantly reduced, as both the 
Kolmogorov-Smirnov and Cram\'er-von Mises tests accept much larger values of the deviations (see Figure~\ref{fig:ecdf_stats}). As a consequence, it is much more
difficult to reject the adequation between any reasonable statistical model and empirical data. We suspect that many GoF tests used in the literature to test models
of financial returns are fundamentally flawed because of the long-ranged volatility correlations. In intuitive terms, the long-memory nature of the volatility can be
thought of as a sequence of volatility regime shifts, each with a different lifetime, and with a broad distribution of these lifetimes. It is clear that in order to
fully sample the unconditional distribution of returns, all the regimes must be encountered several times. In the presence of volatility persistence, therefore, 
the GoF tests are much less stringent, because there is always a possibility that one of these regimes was not, or only partially, sampled. The multifractal model is an
explicit case where this scenario occurs.

We conclude with two remarks of methodological interest. 

1)~The method presented for dealing with self-dependences while using statistical tests of Goodness-of-Fit
is computationally intensive in the sense that it requires to estimate empirically the self-copula for
all lags over the entire unit square. 
In the non-parametric setup, discretization of the space must be chosen so as to provide a good approximation of the continuous distance measures
while at the same time not cause too heavy computations.
Considering that fact, it is often more appropriate to use the Cram\'er-von Mises-like test rather than the Kolmogorov-Smirnov-like, 
as numerical error on the evaluation of the integral will typically be much smaller than on the evaluation of the supremum on a grid,
more so when the grid size is only about $\frac{1}{M}\approx \frac{1}{100}$.

2)~The case with long-ranged dependence $T \gg N \gg 1$ cannot be treated in the framework presented here.
First because the Central Limit Theorem does not hold in that case, and finding the limit law of the statistics may require more advanced mathematics.
But even pre-asymptotically, summing the lags over the available data up to $t\approx N$ means that a lot of noise is included in the determination of $\Psi_N(u,v)$ (see Eqation~\ref{eq:Psi}).
This, in turn, is likely to cause the empirically determined kernel $H(u,v)$ not to be positive definite. One way of addressing this issue is to follow a semi-parametric procedure: 
the copula $C_t$ is still estimated non-parametrically, but the kernel $H$ sums the lagged copulas $C_t$ only up to a scale where the linear correlations and leverage correlations 
vanish, and only one long-ranged dependence mode remains. This last contribution can be fitted by an analytical form, that can then be summed up to its own scale, or even to infinity.

In terms of financial developments, we believe that an empirical exploration of the self-copulas for series of diverse asset returns and at diverse frequencies 
is of primordial importance in order to grasp the complexity of the non-linear time-dependences.
In particular, expanding the concept of the self-copula to pairs of assets is likely to reveal subtle dependence patterns.
From a practitioner's point of view, a multivariate generalization of the self-copula could lead to important progresses on 
such issues as causality, lead-lag effects and the accuracy of multivariate prediction.

\ack
We thank Fr\'ed\'eric Abergel for helpful comments and Vincent Vargas for fruitful discussions.

\newpage 
\appendix

\section{Pseudo-elliptical copula: expansion around independence} 

\begin{table}[b] 
        \caption{Traces of the operators appearing in the covariance functions (multiples of $10^{-2}$). 
                 Traces of the powers of the rank-one $A,R$ equal powers of their traces. The trace of $B+B^{\dagger}$ is 
                 zero. 
                 } 
        \begin{indented} 
        \item[]\begin{tabular}{@{}cccc}\br 
                $I$&$A$&$R$\\\mr 
                16.667&1.176&7.806\\\br 
        \end{tabular} 
        \begin{tabular}{@{}cccc}\br 
                $I^2$&$IA$&$IR$\\\mr 
                111.139&2.948&79.067\\\br 
        \end{tabular} 
        \end{indented} 
        \label{tab:traces} 
\end{table} 

We compute here the spectrum and eigenvectors of the kernel $H(u,v)$ in the case of pseudo-elliptical copula with weak dependences, 
starting from the expansion (\ref{eq:cop_perturb1}). 

The situation is better understood in terms of operators acting in the Hilbert space 
of continuous functions on $[0,1]$ vanishing in the border. 
Using Dirac's braket notations, 
$A=\ket{\tilde{A}}\bra{\tilde{A}},\,B=\ket{\tilde{R}}\bra{\tilde{A}},\,R=\ket{\tilde{R}}\bra{\tilde{R}}$. 
The sine functions $\ket{j}=\sqrt{2}\,\sin(j\pi u)$ build a basis of this Hilbert space, 
and interestingly they are the eigenvectors of the independent kernel $I(u,v)$ 
($I$ stands for `\emph{I}ndependence' and is the covariance matrix of the Brownian motion: 
$I=M-P$ where $M$ denotes the bivariate upper Fr\'echet-Hoeffding copula and $P$ the bivariate product copula). 

It is then easy to find the spectra: rank-one operators have at most one non-null eigenvalue. 
Using the parities of $\tilde{A}(u)$ and $\tilde{R}(u)$ with respect to $\frac{1}{2}$ and imposing orthonormality of the eigenvectors, 
we can sketch the following table of the non zero eigenvalues and eigenvectors of the different operators:

\[
\begin{array}{r@{\,=\,}lp{10pt}cp{10pt}r@{\,=\,}l} 
        \lambda^I_j        &(j\pi)^{-2}                                    &&  U_j^I(u)\,=\,\ket{j}\\ 
        \lambda^R        &\braket{\tilde{R}}{\tilde{R}}={\rm Tr}R  && \ket{U_0^R}\,=\,\ket{\tilde{R}}/\sqrt{{\rm Tr}R}\\ 
        \lambda^A        &\braket{\tilde{A}}{\tilde{A}}={\rm Tr}A  && \ket{U_0^A}\,=\,\ket{\tilde{A}}/\sqrt{{\rm Tr}A}\\ 
\end{array} 
\] 

For the pseudo-elliptical copula with weak dependence, $H$ has the following general form: 
\begin{equation}
	H = I + \tilde\rho R + \tilde\alpha A - \frac{\tilde\beta}{2} (B+B^{\dagger}). 
\end{equation}
The operator $B+B^{\dagger}$ has two non zero eigenvalues $\pm \sqrt{\lambda^R\lambda^A}$, with eigenvectors $(\ket{U_0^R} \pm \ket{U_0^A})/\sqrt{2}$. 
In order to approximately diagonalize $H$, it is useful to notice that in the present context $A$ and $R$ are close to commuting with $I$. 
More precisely, it turns out that $\ket{U_0^A}$ is very close to $\ket{2}$, and $\ket{U_0^R}$ even closer to $\ket{1}$. 
Indeed, \mbox{$a_2=\braket{U_0^A}{2}\approx 0.9934$} and \mbox{$r_1=\braket{U_0^R}{1}\approx 0.9998$}. Using the symmetry of $A$ and $R$, we can therefore 
write: 
\begin{eqnarray*} 
        \ket{U_0^A}&=&a_2\ket{2}+\epsilon_{a}\ket{2_{\perp}}\quad\textrm{with}\quad\braket{2}{2_{\perp}}=\braket{2j\!-\!1}{2_{\perp}}=0,\forall j\geq 1\\ 
        \ket{U_0^R}&=&r_1\ket{1}+\epsilon_{r}\ket{1_{\perp}}\quad\textrm{with}\quad\braket{1}{1_{\perp}}=\braket{2j}{1_{\perp}}=0,\forall j\geq1 
\end{eqnarray*} 
where $\epsilon_a = \sqrt{1-a_2^2} \ll 1$ and $\epsilon_r = \sqrt{1-r_1^2} \ll 1$. The components of $\ket{2_{\perp}}$ on the even eigenvectors of 
$I$ are determined as: 
\[ 
\braket{2_{\perp}}{2j} = \frac{\braket{U_0^A}{2j}}{\epsilon_a} \qquad j \geq 2, 
\] 
and similarly: 
\[ 
\braket{1_{\perp}}{2j\!-\!1} = \frac{\braket{U_0^R}{2j\!-\!1}}{\epsilon_r} \qquad j \geq 2. 
\] 

Using the definition of the coefficients $\alpha_t, \beta_t$ and $\rho_t$ given in section 3, we introduce the following notations: 
\begin{eqnarray*} 
        \tilde\alpha&=&2\,{\rm Tr}A\lim_{N\to\infty}\sum_{t=1}^{N-1}\left(1-\frac{t}{N}\right)\alpha_t\\ 
        \tilde\rho  &=&2\,{\rm Tr}R\lim_{N\to\infty}\sum_{t=1}^{N-1}\left(1-\frac{t}{N}\right)\rho_t\\ 
        \tilde\beta &=&2\,\sqrt{{\rm Tr}A\,{\rm Tr}R}\lim_{N\to\infty}\sum_{t=1}^{N-1}\left(1-\frac{t}{N}\right)\beta_t 
\end{eqnarray*} 
so that $H$ writes: 
\begin{eqnarray*} 
        H&=I &+\tilde\alpha\ket{U_0^A}\bra{U_0^A}+\tilde\rho\ket{U_0^R}\bra{U_0^R}-\tilde\beta\overleftrightarrow{\ket{U_0^R}\bra{U_0^A}}\\ 
         &=H_0
          &+\epsilon_a \left( \tilde\alpha a_2 \overleftrightarrow{\ket{2}\bra{2_{\perp}}}-\tilde\beta r_1a_{\perp}\overleftrightarrow{\ket{1}\bra{2_{\perp}}}\right)\\ 
         &&+\epsilon_r \left( \tilde\rho r_1 \overleftrightarrow{\ket{1_{\perp}}\bra{1}}-\tilde\beta a_2\overleftrightarrow{\ket{1_{\perp}}\bra{2}}\right)\\ 
         &&+\left(\tilde\alpha \epsilon_a^2\ket{2_{\perp}}\bra{2_{\perp}}+\tilde\rho \epsilon_r^2 \ket{1_{\perp}}\bra{1_{\perp}}-\tilde\beta \epsilon_a \epsilon_r\overleftrightarrow{\ket{2_{\perp}}\bra{1_{\perp}}}\right) 
\end{eqnarray*} 
where $\overleftrightarrow{\ket{\psi_1}\bra{\psi_2}}=\frac{1}{2}\big(\ket{\psi_1}\bra{\psi_2}+\ket{\psi_2}\bra{\psi_1} \big)$
and $H_0$ is the unperturbed operator (0-th order in both $\epsilon$s)
\[
	H_0=\sum_{j\geq 3}\lambda_j^I\ket{j}\bra{j}+(\lambda_2^I+\tilde\alpha a_2^2)\ket{2}\bra{2}+(\lambda_1^I+\tilde\rho r_1^2)\ket{1}\bra{1}-\tilde\beta r_1a_2\overleftrightarrow{\ket{2}\bra{1}}
\]
the spectrum of which is easy to determine as: 
\[ 
\begin{array}{r@{=}lp{10pt}r@{=}l} 
        \lambda_1^{H_0}&\lambda_-\xrightarrow{\tilde\rho,\tilde\beta\to 0}{\lambda_1^I}&& 
        \ket{U_1^{H_0}}&-\displaystyle\frac{\ket{-}}{\sqrt{\braket{-}{-}}} \xrightarrow{\tilde\rho,\tilde\beta\to 0}{\ket{1}}\\ 
        \lambda_2^{H_0}&\lambda_+\xrightarrow{\tilde\rho,\tilde\beta\to 0}{\lambda_2^I+\tilde\alpha a_2^2}&& 
        \ket{U_2^{H_0}}&\phantom{-}\displaystyle \frac{\ket{+}}{\sqrt{\braket{+}{+}}} \xrightarrow{\tilde\rho,\tilde\beta\to 0}{\ket{2}}\\ 
        \lambda_j^{H_0}&\lambda_j^I&& 
        \ket{U_j^{H_0}}& \ket{j} \qquad (j \geq 3) 
\end{array} 
\] 
where 
\[ 
\lambda_{\pm}=\frac{\lambda_1^I+\tilde\rho r_1^2+\lambda_2^I+\tilde\alpha a_2^2\pm\sqrt{(\lambda_1^I+\tilde\rho r_1^2-\lambda_2^I-\tilde\alpha a_2^2)^2+4(\tilde\beta r_1a_2)^2}}{2} 
\] 
and $\ket{\pm}$
the corresponding eigenvectors, which are linear combination of $\ket{1}$ and $\ket{2}$ only. Therefore, $\braket{1_{\perp}}{\pm}=\braket{2_{\perp}}{\pm}=0$. This implies that there is no corrections to the eigenvalues of $H_0$ to first order in the $\epsilon$s. 

At the next order, instead, some corrections appear. We call: 
\begin{eqnarray*}
	V_{i,j}&=(\tilde\rho r_1\braket{1}{U_i^{H_0}}-\frac{\tilde\beta a_2}{2}\braket{2}{U_i^{H_0}})\braket{j}{1_{\perp}}\epsilon_r\\
	       &+(\tilde\alpha a_2\braket{2}{U_i^{H_0}}-\frac{\tilde\beta r_1}{2} \braket{1}{U_i^{H_0}})\braket{j}{2_{\perp}}\epsilon_a 
\end{eqnarray*} 
the matrix elements of the first order perturbation of $H$, whence 
\begin{eqnarray*} 
        \lambda_1^{H}&=\lambda_1^{H_0}+\sum_{j\geq 3}\frac{V_{1,j}^2}{\lambda_1^{H_0}-\lambda_j^{H_0}}\\ 
        \lambda_2^{H}&=\lambda_2^{H_0}+\sum_{j\geq 3}\frac{V_{2,j}^2}{\lambda_2^{H_0}-\lambda_j^{H_0}}\\ 
        \lambda_j^{H}&         =\lambda_j^{H_0} +\sum_{i=1,2}\frac{V_{i,j}^2}{\lambda_j^{H_0}-\lambda_i^{H_0}} 
                                                 +\big(\tilde\alpha \epsilon_a^2\braket{j}{2_{\perp}}^2+\tilde\rho \epsilon_r^2\braket{j}{1_{\perp}}^2-\tilde\beta \epsilon_a \epsilon_r \braket{j}{1_{\perp}}\braket{j}{2_{\perp}}\big) 
\end{eqnarray*} 
As of the eigenvectors, it is enough to go to first order in $\epsilon$s to get a non-trivial perturbative correction: 
\begin{eqnarray*} 
        \ket{U_1^H}&=&\ket{U_1^{H_0}}+\sum_{j\geq 3}\frac{V_{1,j}}{\lambda_1^{H_0}-\lambda_j^{H_0}}\ket{j}\\ 
        \ket{U_2^H}&=&\ket{U_2^{H_0}}+\sum_{j\geq 3}\frac{V_{2,j}}{\lambda_2^{H_0}-\lambda_j^{H_0}}\ket{j}\\ 
        \ket{U_j^H}&=&\ket{j}        +\sum_{i=1,2  }\frac{V_{i,j}}{\lambda_j^{H_0}-\lambda_i^{H_0}}\ket{U_i^{H_0}} 
\end{eqnarray*}

\paragraph{}
The special case treated numerically in section 3 corresponds to $\tilde\rho =\tilde\beta = 0$, such that the above expressions simplify 
considerably, since in that case $V_{1,j} \equiv 0$ and $V_{2,2j-1}=0$, while $V_{2,2j} = \tilde\alpha a_2 \braket{U_0^A}{2j}$.
To first order in the $\epsilon$s, the spectrum is not perturbed and calls $\lambda_i^{H}=\lambda_i^{H_0}=\lambda_i^{I}+\tilde\alpha a_2^2\delta_{i2}$, so
that the characteristic function of the modified CM distribution is, according to Equation~(\ref{eq:charctCM}),
\[
	\phi(t)=\prod_{j}\left(1-2\mathrm{i}t/(j\pi)^2\right)^{-\frac{1}{2}} \times \sqrt{\frac{1-2\mathrm{i}t\lambda_2^{I}}{1-2\mathrm{i}t\lambda_2^{H_0}}}.
\]
Its pdf is thus the convolution of the Fourier transform of $\phi_I(t)$ (characteristic function associated to the usual CM distribution \cite{anderson1952asymptotic})
and the Fourier transform of the correction $\phi_c(t)=\sqrt{1-2\mathrm{i}t\lambda_2^{I}}/\sqrt{1-2\mathrm{i}t\lambda_2^{H_0}}$.
Noting that $(1-2\mathrm{i}\sigma^2t)^{-\frac{1}{2}}$ is the characteristic function of the chi-2 distribution, it can be shown that for $k > 0$, and
with $\mu\equiv\lambda_2^{H_0}$ for the sake of readability:
\begin{eqnarray*}
	\frac{1}{\sqrt{2\pi}}{\rm FT}\left(\phi_c\right)&=\delta(k)-
	\int_{\lambda_2^{I}}^{\mu}{\rm d}\lambda\frac{\partial}{\partial k} \left(\chi^2(k;\mu)*\chi^2(k;\lambda)\right)\\
	&=\delta(k)-\int_{\lambda_2^{I}}^{\mu}{\rm d}\lambda\frac{\e^{-\frac{\lambda+\mu}{4\lambda\mu}k}}{8(\lambda\mu)^{\frac{3}{2}}}
	  \left((\mu-\lambda)I_1(\frac{\mu-\lambda}{4\lambda\mu}k)-(\mu+\lambda)I_0(\frac{\mu-\lambda}{4\lambda\mu}k)\right)\\
	&\approx\delta(k)+\e^{-\frac{k}{2\lambda}}\frac{\tilde\alpha a_2^2}{4\lambda^2} I_0(\frac{\tilde\alpha a_2^2}{4\lambda^2}k)\\
\end{eqnarray*}
where $\chi^2(k;\sigma^2)=(2\pi\sigma^2k\,\e^{k/\sigma^2})^{-\frac{1}{2}}$ is the pdf of the chi-2 distribution, 
$I_{n}$ are the modified Bessel functions of the first kind, and $*$ denotes the convolution operation.
The approximation on the last line holds as long as $\tilde\alpha\ll\lambda_2^I=(2\pi)^{-2}$ and in this regime we obtain finally
\begin{eqnarray*}
	\mathcal{P}[C\!M=k]&=\sqrt{2\pi}{\rm FT}(\phi)(k)=({\rm FT}(\phi_I)*{\rm FT}(\phi_c))(k)\\
	                 &=\mathcal{P}_I(k)+{4\tilde\alpha a_2^2\pi^4}\int_{0}^{k}\mathcal{P}_I(z)\e^{-2\pi^2{(k-z)}} I_0(4\tilde\alpha a_2^2\pi^4(k-z)){\rm d}z\\
	                 &=\mathcal{P}_I(k)+{4\tilde\alpha a_2^2\pi^4}\int_{0}^k\mathcal{P}_I(k-z)\e^{-2\pi^2{z}} I_0(4\tilde\alpha a_2^2\pi^4z){\rm d}z
\end{eqnarray*}

\begin{figure}[b]
	\begin{indented}
	\item[]
	\includegraphics[scale=0.5]{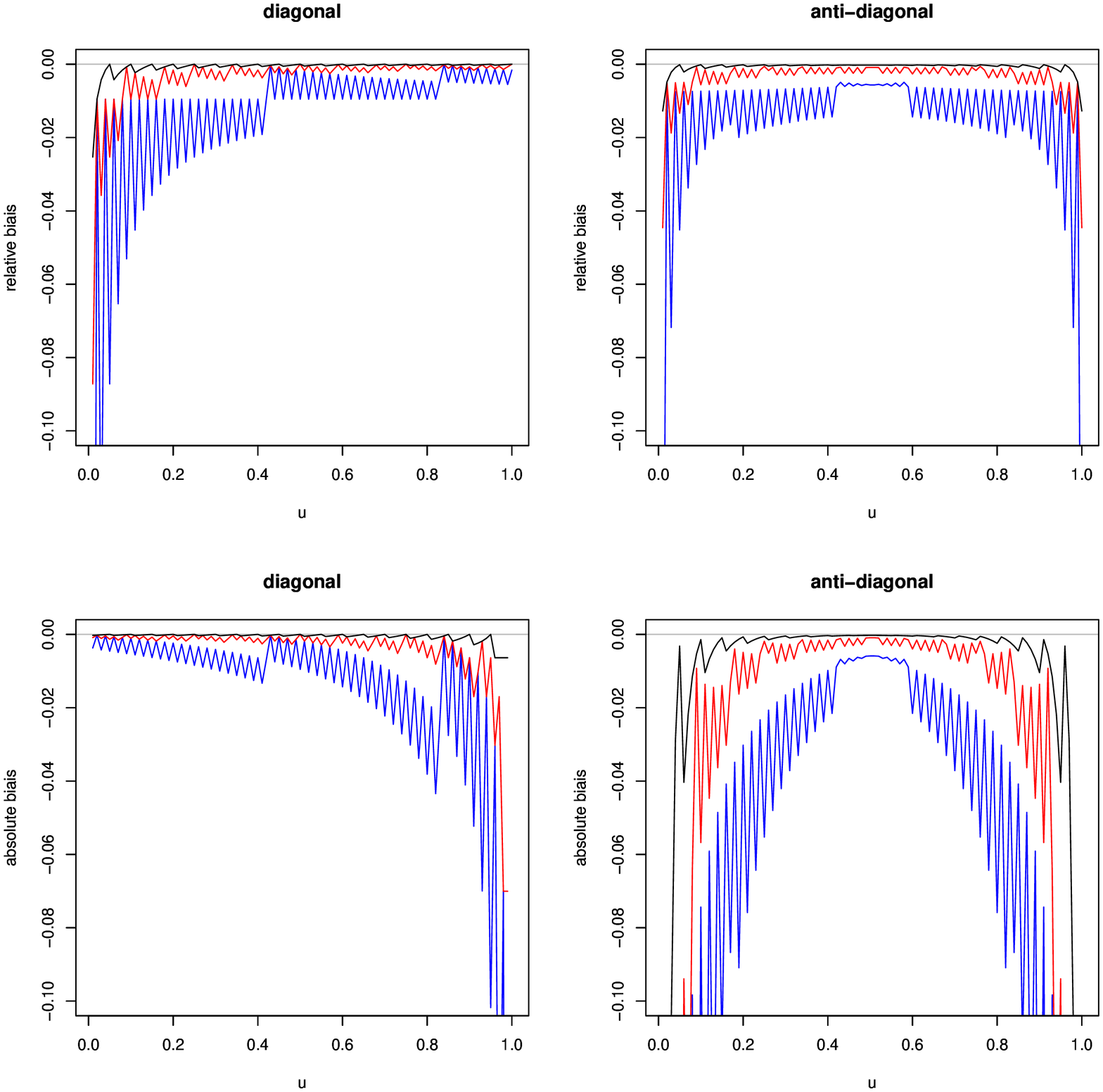}
	\end{indented}
	\caption{Bias of the non-parametric naive estimator of the independence copula with unknown marginals, 
	         for different sample sizes $N=251$ (blue), $1256$ (red), $6280$ (black).
	\textbf{Top:} Relative bias $\frac{\lfloor Nu\rfloor}{Nu}\frac{\lfloor Nv\rfloor}{Nv}-1$.
	\textbf{Bottom:} Absolute bias $uv\left(\frac{\lfloor Nu\rfloor}{Nu}\frac{\lfloor Nv\rfloor}{Nv}-1\right)/(\min(u,v)-uv)$}
	\label{fig:bias}
\end{figure}

\newpage
\section{Non-parametric copula estimator} 

The copula $C(u,v)$ of a random pair $(X,Y)$ is
\[
	C(u,v)=\mathds{P}[F_X(X)\leq u, F_Y(Y)\leq v]=\mathds{E}\left[\mathds{1}_{\{X\leq F^{-1}_X(u)\}}\mathds{1}_{\{Y\leq F^{-1}_Y(v)\}}\right]
\]
If the univariate marginals $F_X,F_Y$ are known, the usual empirical counterpart to the expectation operator can be used to define the empirical copula
over a sample of $N$ i.d. realisations $(X_i,Y_i)$ of the random pair:
\[
	\overline{C}(u,v)=\frac{1}{N}\sum_{i=1}^N\mathds{1}_{\{X_i\leq F^{-1}_X(u)\}}\mathds{1}_{\{Y_i\leq F^{-1}_Y(v)\}}
\]
which is clearly unbiased for any $N$.
But if the marginals are unknown, they have to be themselves estimated, for example by their usual empirical counterpart. 
Since $F(x)=\mathds{P}[X\leq x]=\mathds{E}[\mathds{1}_{\{X\leq x\}}]$, an unbiased ecdf is obtained as
$\overline{F}(x)=\frac{1}{N}\sum_{i=1}^N\mathds{1}_{\{X_i\leq x\}}$, but the expectation value of
\[
	\overline{C}(u,v)=\frac{1}{N}\sum_{i=1}^N\mathds{1}_{\{\overline{F}_X(X_i)\leq u\}}\mathds{1}_{\{\overline{F}_Y(Y_i)\leq v\}}
\]
is not $C(u,v)$ anymore (only asymptotically is $\overline{C}(u,v)$ unbiased), but rather
\[
	\mathds{E}[\overline{C}(u,v)]=\int dF_{XY}(x,y)\mathds{P}\left[B(x)\leq Nu-1,B(y)\leq Nv-1\right]
\]
where $B_X(x)=\sum_{j<N}\mathds{1}_{\{X_j\leq x\}}$ has a binomial distribution $\mathcal{B}(p,N-1)$ with $p=F(x)$
and is not independent of $B_Y(y)$.
As an example, the expected value of the independence (product) copula $C(u,v)=uv$ is 
\[
	\mathds{E}[\overline{C}(u,v)]=\frac{\lfloor Nu\rfloor}{N}\frac{\lfloor Nv\rfloor}{N}\equiv n(u,v) uv
\]
resulting in a relative bias $n(u,v)-1=\frac{\lfloor Nu\rfloor}{Nu}\frac{\lfloor Nv\rfloor}{Nv}-1$ vanishing only asymptotically.

As $\mathds{E}[\overline{C}(u,v)]$ may not be computable in the general case, we define as 
\[
	\frac{Nu}{\lfloor Nu\rfloor}\frac{Nv}{\lfloor Nv\rfloor}\,\overline{C}(u,v)
\]
our non-parametrical estimator of the copula with bias correction, even when $C$ is not the independence copula.
Therefore, our estimator is technically biased at finite $N$ but with a good bias correction, and asymptotically unbiased.

\newpage
\nocite{weiss1978modification}
\bibliographystyle{ieeetr}
\bibliography{../biblio_all}

\begin{thebibliography}{10}

\bibitem{darling1957kolmogorov}
D.~A. Darling, ``{The {K}olmogorov-{S}mirnov, {C}ramer-von {M}ises Tests},''
  {\em The Annals of Mathematical Statistics}, vol.~28, no.~4, pp.~823--838,
  1957.

\bibitem{plerou1999scaling}
V.~Plerou, P.~Gopikrishnan, L.~A.~N. Amaral, M.~Meyer, and H.~E. Stanley,
  ``Scaling of the distribution of price fluctuations of individual
  companies,'' {\em Physical Review E}, vol.~60, no.~6, p.~6519, 1999.

\bibitem{dragulescu2002probability}
A.~A. Dragulescu and V.~M. Yakovenko, ``Probability distribution of returns in
  the heston model with stochastic volatility,'' {\em Quantitative Finance},
  vol.~2, no.~6, pp.~443--453, 2002.

\bibitem{bouchaud2003theory}
J.-P. Bouchaud and M.~Potters, {\em {Theory of Financial Risk and Derivative
  Pricing: from Statistical Physics to Risk Management}}.
\newblock Cambridge University Press, 2003.

\bibitem{malevergne2006extreme}
Y.~Malevergne and D.~Sornette, {\em {Extreme financial risks: From dependence
  to risk management}}.
\newblock Springer Verlag, 2006.

\bibitem{embrechts2003modelling_art}
P.~Embrechts, F.~Lindskog, and A.~McNeil, ``Modelling dependence with copulas
  and applications to risk management,'' {\em Handbook of heavy tailed
  distributions in finance}, vol.~8, pp.~329--384, 2003.

\bibitem{embrechts2002correlation_art}
P.~Embrechts, A.~McNeil, and D.~Straumann, ``Correlation and dependence in risk
  management: properties and pitfalls,'' {\em Risk management: value at risk
  and beyond}, pp.~176--223, 2002.

\bibitem{mikosch2006copulas}
T.~Mikosch, ``{Copulas: Tales and facts},'' {\em Extremes}, vol.~9, no.~1,
  pp.~3--20, 2006.

\bibitem{chicheportiche2010joint}
R.~Chicheportiche and J.-P. Bouchaud, ``{The joint distribution of stock
  returns is not elliptical},'' {\em Arxiv preprint q-fin.ST/1009.1100}, 2010.

\bibitem{beare2010copulas}
B.~K. Beare, ``Copulas and temporal dependence,'' {\em Econometrica}, vol.~78,
  no.~1, pp.~395--410, 2010.

\bibitem{ibragimov2008copulas}
R.~Ibragimov and G.~Lentzas, ``Copulas and long memory.'' Harvard Institute of
  Economic Research discussion paper, 2008.

\bibitem{patton2009copula_art}
A.~J. Patton, ``Copula-based models for financial time series,'' {\em Handbook
  of financial time series}, pp.~767--785, 2009.

\bibitem{darsow1992copulas}
W.~F. Darsow, B.~Nguyen, and E.~T. Olsen, ``Copulas and {M}arkov processes,''
  {\em Illinois Journal of Mathematics}, vol.~36, no.~4, pp.~600--642, 1992.

\bibitem{bradley2007introduction}
R.~C. Bradley, {\em Introduction to strong mixing conditions}, vol.~1--3.
\newblock Kendrick Press, Heber City, Utah, 2007.

\bibitem{chen2010nonlinearity}
X.~Chen, L.~P. Hansen, and M.~Carrasco, ``Nonlinearity and temporal
  dependence,'' {\em Journal of Econometrics}, vol.~155, no.~2, pp.~155--169,
  2010.

\bibitem{anderson1952asymptotic}
T.~W. Anderson and D.~A. Darling, ``{Asymptotic Theory of Certain ``{G}oodness
  of {F}it" Criteria Based on Stochastic Processes},'' {\em The Annals of
  Mathematical Statistics}, vol.~23, no.~2, pp.~193--212, 1952.

\bibitem{mandelbrot1968fractional}
B.~B. Mandelbrot and J.~W. Van~Ness, ``Fractional brownian motions, fractional
  noises and applications,'' {\em SIAM review}, vol.~10, no.~4, pp.~422--437,
  1968.

\bibitem{avramov2006liquidity}
D.~Avramov, T.~Chordia, and A.~Goyal, ``Liquidity and autocorrelations in
  individual stock returns,'' {\em The Journal of Finance}, vol.~61, no.~5,
  pp.~2365--2394, 2006.

\bibitem{pasquini1999multiscaling}
M.~Pasquini and M.~Serva, ``Multiscaling and clustering of volatility,'' {\em
  Physica A: Statistical Mechanics and its Applications}, vol.~269, no.~1,
  pp.~140--147, 1999.

\bibitem{calvetfisher}
L.~E. Calvet and A.~Fisher, {\em {Multifractal Volatility: Theory, Forecasting,
  and Pricing}}.
\newblock Academic Press, 2008.

\bibitem{bacrymuzy}
J.-F. Muzy, E.~Bacry, and J.~Delour, ``{Modelling fluctuations of financial
  time series: from cascade process to stochastic volatility model},'' {\em The
  European Physical Journal B}, vol.~17, no.~3, pp.~537--548, 2000.

\bibitem{zumbach2001heterogeneous}
G.~Zumbach and P.~Lynch, ``Heterogeneous volatility cascade in financial
  markets,'' {\em Physica A: Statistical Mechanics and its Applications},
  vol.~298, no.~3-4, pp.~521--529, 2001.

\bibitem{lynch2003market}
P.~Lynch and G.~Zumbach, ``Market heterogeneities and the causal structure of
  volatility,'' {\em Quantitative Finance}, vol.~3, no.~4, pp.~320--331, 2003.

\bibitem{borland2005dynamics}
L.~Borland, J.-P. Bouchaud, J.-F. Muzy, and G.~Zumbach, ``The dynamics of
  financial markets --- {M}andelbrot's multifractal cascades, and beyond,''
  {\em Wilmott Magazine}, pp.~86--96, March 2005.

\bibitem{bouchaud2001more}
J.-P. Bouchaud and M.~Potters, ``More stylized facts of financial markets:
  leverage effect and downside correlations,'' {\em Physica A: Statistical
  Mechanics and its Applications}, vol.~299, no.~1-2, pp.~60--70, 2001.

\bibitem{perello2004multiple}
J.~Perell{\'o}, J.~Masoliver, and J.-P. Bouchaud, ``Multiple time scales in
  volatility and leverage correlations: a stochastic volatility model,'' {\em
  Applied Mathematical Finance}, vol.~11, no.~1, pp.~27--50, 2004.

\bibitem{pochart2002skewed}
B.~Pochart and J.-P. Bouchaud, ``The skewed multifractal random walk with
  applications to option smiles,'' {\em Quantitative Finance}, vol.~2, no.~4,
  pp.~303--314, 2002.

\bibitem{eisler2004multifractal}
Z.~Eisler and J.~Kertesz, ``Multifractal model of asset returns with leverage
  effect,'' {\em Physica A: Statistical Mechanics and its Applications},
  vol.~343, pp.~603--622, 2004.

\bibitem{ahlgren2007frustration}
P.~T. Ahlgren, M.~H. Jensen, I.~Simonsen, R.~Donangelo, and K.~Sneppen,
  ``Frustration driven stock market dynamics: Leverage effect and asymmetry,''
  {\em Physica A: Statistical Mechanics and its Applications}, vol.~383, no.~1,
  pp.~1--4, 2007.

\bibitem{reigneron2011principal}
P.-A. Reigneron, R.~Allez, and J.-P. Bouchaud, ``Principal regression analysis
  and the index leverage effect,'' {\em Physica A: Statistical Mechanics and
  its Applications}, 2011.

\bibitem{blomqvist1950measure}
N.~Blomqvist, ``{On a measure of dependence between two random variables},''
  {\em The Annals of Mathematical Statistics}, vol.~21, no.~4, pp.~593--600,
  1950.

\bibitem{muzy2001multifractal}
J.-F. Muzy, D.~Sornette, J.~Delour, and A.~Arneodo, ``{Multifractal returns and
  hierarchical portfolio theory},'' {\em Quantitative Finance}, vol.~1, no.~1,
  pp.~131--148, 2001.

\bibitem{lux}
T.~Lux, ``{The Multi-Fractal Model of Asset Returns: Its Estimation via {GMM}
  and Its Use for Volatility Forecasting},'' {\em Journal of Business and
  Economic Statistics}, vol.~26, p.~194, 2008.

\bibitem{duchon2010forecasting}
J.~Duchon, R.~Robert, and V.~Vargas, ``Forecasting volatility with the
  multifractal random walk model,'' {\em Mathematical Finance}, 2008.

\bibitem{borland2005multi}
L.~Borland and J.-P. Bouchaud, ``On a multi-timescale statistical feedback
  model for volatility fluctuations,'' {\em Arxiv preprint
  physics.soc-ph/0507073}, 2005.

\bibitem{muzy2006extreme}
J.-F. Muzy, E.~Bacry, and A.~Kozhemyak, ``Extreme values and fat tails of
  multifractal fluctuations,'' {\em Physical Review E}, vol.~73, no.~6,
  p.~066114, 2006.

\bibitem{bacry2006asset}
E.~Bacry, A.~Kozhemyak, and J.-F. Muzy, ``Are asset return tail estimations
  related to volatility long-range correlations?,'' {\em Physica A: Statistical
  Mechanics and its Applications}, vol.~370, no.~1, pp.~119--126, 2006.

\bibitem{weiss1978modification}
M.~S. Weiss, ``{Modification of the {K}olmogorov-{S}mirnov statistic for use
  with correlated data},'' {\em Journal of the American Statistical
  Association}, vol.~73, no.~364, pp.~872--875, 1978.

\end{thebibliography}

\end{document}